\documentclass[12pt]{article}
\usepackage{graphicx,amsmath,amssymb,cite}

\newlength{\dinwidth}
\newlength{\dinmargin}
\begin{document}
\titlepage
\begin{flushright}
  IPPP/10/29 \\
  DCPT/10/58 \\
  Cavendish-HEP-10/07 \\
  CERN-PH-TH/2010-160 \\
  22nd November 2010 \\
\end{flushright}

\vspace*{0.5cm}

\begin{center}
  {\Large \bf Heavy-quark mass dependence in global PDF analyses\\[1ex] and 3- and 4-flavour parton distributions}
  
  \vspace*{1cm}
  \textsc{A.D. Martin$^a$, W.J. Stirling$^b$, R.S. Thorne$^c$ and G. Watt$^d$} \\
  
  \vspace*{0.5cm}$^a$ Institute for Particle Physics Phenomenology, University of Durham, DH1 3LE, UK \\
  $^b$ Cavendish Laboratory, University of Cambridge, CB3 0HE, UK \\
  $^c$ Department of Physics and Astronomy, University College London, WC1E 6BT, UK \\
  $^d$ Theory Group, Physics Department, CERN, CH-1211 Geneva 23, Switzerland
\end{center}

\vspace*{0.5cm}

\begin{abstract}
  We study the sensitivity of our recent MSTW~2008 NLO and NNLO PDF analyses to the values of the charm- and bottom-quark masses, and we provide additional public PDF sets for a wide range of these heavy-quark masses.  We quantify the impact of varying $m_c$ and $m_b$ on the cross sections for $W$, $Z$ and Higgs production at the Tevatron and the LHC.  We generate 3- and 4-flavour versions of the (5-flavour) MSTW~2008 PDFs by evolving the input PDFs and $\alpha_S$ determined from fits in the 5-flavour scheme, including the eigenvector PDF sets necessary for calculation of PDF uncertainties.  As an example of their use, we study the difference in the $Z$ total cross sections at the Tevatron and LHC in the 4- and 5-flavour schemes.  Significant differences are found, illustrating the need to resum large logarithms in $Q^2/m_b^2$ by using the 5-flavour scheme.  The 4-flavour scheme is still necessary, however, if cuts are imposed on associated (massive) $b$-quarks, as is the case for the experimental measurement of $Zb\bar{b}$ production and similar processes.
\end{abstract}

\vspace*{0.5cm}

\section{Introduction} \label{sec:introduction}
Parton distribution functions (PDFs) determined from global analysis~\cite{Martin:2009iq,Martin:2007bv,Lai:2010vv,Nadolsky:2008zw,Ball:2010de} of fixed-target and collider data, mainly from deep-inelastic scattering (DIS), are an essential input to theory calculations for hard processes at hadron colliders such as the Tevatron and LHC.  In addition to the fitted input PDFs, several other parameters enter into the global fits, which affect both the PDFs obtained and predictions for hadronic processes.  One important example is the value of the strong coupling $\alpha_S$ and its uncertainty~\cite{Martin:2009bu,Lai:2010nw,Demartin:2010er}.  Others, which are the focus of the present paper, are the values of the heavy-quark masses and the scheme used to calculate heavy-quark contributions to observables.  In particular, while the precise values taken for $m_c$ and $m_b$ may not be crucial for the processes included in the global PDF analyses, calculations of processes with explicit $c$- and $b$-quarks might be more sensitive to these values, and it is therefore desirable to have PDFs available which have been consistently fitted and evolved with corresponding values of $m_c$ and $m_b$.

We will study two topics, both of which concern the treatment of the heavy charm and bottom quarks in global parton analyses.  The first topic, which is the subject of Section~\ref{sec:massdep}, concerns the sensitivity of the MSTW~2008 global parton analysis~\cite{Martin:2009iq} to the values of the heavy-quark masses $m_h$, with $h=c,b$.  In Ref.~\cite{Martin:2009iq} these masses were taken to be $m_c=1.40$~GeV and $m_b=4.75$~GeV.  Here we perform global fits for a range of $m_c$ and $m_b$ about these values, with $\alpha_S(M_Z^2)$ allowed to be a free parameter.  In this way, we determine the values of $m_c$ and $m_b$ preferred by the data, and the correlations between these values of $m_h$ and $\alpha_S(M_Z^2)$.  Due to a significant correlation between $m_c$ and $\alpha_S(M_Z^2)$, we also perform fits with varying $m_c$ but with $\alpha_S(M_Z^2)$ held fixed.  We study how the cross sections for $W$, $Z$ and Higgs boson production at the Tevatron and the LHC depend on the choice of $m_c$ and $m_b$, and we provide a recommendation on how to include the uncertainty arising from $m_h$ in a general cross-section calculation.

The second topic, described in Section~\ref{sec:FFNS}, is the generation of sets of 3- and 4-flavour PDFs corresponding to the MSTW~2008~\cite{Martin:2009iq} 5-flavour sets of PDFs.  We follow the same procedure previously used to generate the 3- and 4-flavour versions~\cite{Martin:2006qz} of the 5-flavour MRST 2004 PDFs~\cite{Martin:2004ir}, i.e.~the input PDFs (and $\alpha_S$) at $Q_0^2=1$~GeV$^2$ determined from fits in the 5-flavour scheme are used as the initial condition for 3- or 4-flavour evolution.  However, going beyond Ref.~\cite{Martin:2006qz}, we also provide 3- and 4-flavour \emph{eigenvector} PDF sets to enable calculation of PDF uncertainties, and we provide 3- and 4-flavour PDF sets for a wide range of $m_c$ and $m_b$ values, respectively.  As an example of the use of the central 4-flavour PDFs for the default quark masses, we compare the total cross sections for $Z$ production at the Tevatron and LHC in the 4- and 5-flavour schemes.  However, first we begin with a short r\'{e}sum\'{e} of the alternative ``schemes'' to treat heavy flavours\footnote{Strictly speaking, it would be better to say alternative ``techniques'', since the use of the word ``scheme'' is usually reserved for an alternative choice in the ordering of the perturbative expansion, or a particular separation of contributions between the coefficient functions and parton densities---an ambiguity inherent in QCD.}, and in particular, an explanation of what precisely we mean by 3-, 4-, and 5-flavour PDFs.

\section{Schemes for the treatment of heavy flavours}
It is appropriate to briefly recall the various schemes for the treatment of heavy flavours in global parton analyses.  In PDF analyses it is common to start evolving upwards from $Q^2=Q^2_0\sim 1$~GeV$^2$ with the distributions of the three light quarks $(u,d,s)$, assuming that they are massless.  As we evolve upwards, we have the choice to turn on the heavy-quark $(c,b,t)$ distributions as we pass through their respective transition point, for which a convenient choice is $\mu_F^2=m_h^2$.  As we pass through a transition point, the number of active quarks is increased from $n$ to $n+1$, and the parton densities are related to each other perturbatively, i.e.
\begin{equation}
  f^{n+1}_j(\mu_F^2)=\sum_k A_{jk}(\mu_F^2/m_h^2)\otimes f^{n}_k(\mu_F^2),
  \label{eq:transition}
\end{equation}
 where the perturbative matrix elements $A_{jk}(\mu_F^2/m_h^2)$ contain $\ln(\mu_F^2/m_h^2)$ terms which are known to ${\cal O} (\alpha_S^2)$~\cite{Buza:1996wv} and ${\cal O} (\alpha_S^3)$~\cite{Bierenbaum:2009mv}. The ``$x$'' arguments have been suppressed in Eq.~(\ref{eq:transition}), and the symbol $\otimes$ is shorthand for the convolution
\begin{equation}
  f \otimes g \equiv \int^1_x \frac{{\rm d}x'}{x'}f(x')g(x/x').
\end{equation}
Eq.~(\ref{eq:transition}) relates the $f^{n}_i(\mu_F^2)$ set of partons to the $f^{n+1}_i(\mu_F^2)$ set, guaranteeing the correct evolution for both the $n$ and $n+1$ regimes.  We make the simplest choice, $\mu_F^2=Q^2$, for the factorisation scale.

Hence, we have to decide whether or not to keep a heavy quark as just a final-state particle, and not as a parton within the proton.  We may choose to keep just the three light flavours as parton distributions.  We will call this the 3-flavour scheme (3FS), though it is often referred to as the fixed flavour number scheme (FFNS).  Alternatively, we may include the $c$-quark in the evolution above $Q^2=m_c^2$ and generate 4-flavour PDFs in a 4-flavour scheme (4FS).  Actually, in the global MRST/MSTW parton analyses we also include the $b$-quark distribution in the evolution above $Q^2=m_b^2$, but \emph{not} the $t$-quark above $Q^2=m_t^2$, so we generate 5-flavour sets of parton distributions in a 5-flavour scheme (5FS).  So to be precise, in our definition of $n_f$-flavour parton sets, $n_f$ refers to the {\it maximum} number of quark flavours in the evolution.

In each $n_f$-flavour scheme ($n_f$FS) the structure functions are given by the usual convolution of coefficient functions and parton distributions:
\begin{equation}
  F(x,Q^2)=\sum_{j} C^{n_f\rm FS}_j(Q^2/m_h^2)\otimes f^n_j(Q^2),
  \label{general}
\end{equation}
where the sum $j$ is over the gluon and the (variable, depending on $Q^2$) number of active quark flavours, $n\le n_f$.  We have a choice in how to choose $n_f$ and define the coefficient functions.  One simple choice is to fix $n=n_f=3$.  For the heavy flavours, all the $m_h$ dependence then occurs in the coefficient functions, and these are called the FFNS coefficient functions.  The structure functions may be written in the form
\begin{equation}
  F(x,Q^2)=\sum_{j=u,d,s,g} C^{\rm FF,3}_j(Q^2/m_h^2)\otimes f^3_j(Q^2).
  \label{ffns}
\end{equation}
However, the sum of the $\alpha_S^k\ln^m(Q^2/m_h^2)$ terms, with $m\leq k$, is not included in the perturbative expansion.  Thus the accuracy of the expansion becomes increasingly unreliable as $Q^2$ increases above $m_h^2$.  In addition, there is the problem that the full mass dependence of the coefficient functions is known up to NLO~\cite{Laenen:1992zk}, but is not completely defined at NNLO, i.e.~the $\alpha_S^3$ coefficient, $C^{{\rm FF},3,(3)}_{2, hg}$, for $F_2$ is not fully known, see~Ref.~\cite{Bierenbaum:2009mv}.  (Here, the outer subscript ``$hg$'' denotes the $g\to h$ contribution to the heavy-flavour structure function $F_2^h$.)

As an aside, we note that it would be possible to treat the charm quark as light, and the bottom quark as heavy.  Then it would be possible to express the structure functions and cross sections in terms of four light quarks with all the mass dependence of the bottom quark contained in the coefficient functions.  This is sometimes called the 4-flavour FFNS.  However if one needs to use scales $Q^2<m^2_b$ and in the vicinity of $m_c^2$, as for example in the global PDF analysis, then the charm-quark mass dependence, and the charm transition point (if $Q^2 \leq m_c^2$), should be included.  Thus, this 4-flavour FFNS is only applicable in a restricted range of $Q^2$, and it will not be considered further here.

The alternative, and better, approach is to use the 4-flavour (and 5-flavour) PDFs of the variable flavour number schemes (4FS, 5FS).  The simplest variable flavour number evolution procedure is to evolve treating some (or all) of the heavy quarks as massless, but to turn on the distributions at the appropriate transition points, $Q^2=m_h^2$.  That is, to assume that these heavy-quark distributions evolve according to the splitting functions for massless quarks.  Thus, the resummation of the large logarithms in $Q^2/m_h^2$ is achieved by the introduction of heavy-flavour parton distributions and the solution of the evolution equations.  It is motivated by the observation that at high scales the massive quarks behave like massless partons, and the coefficient functions are simply those in the massless limit, e.g.~for structure functions
\begin{equation}
  F(x,Q^2) =\sum_j C^{n_f{\rm ZMVF}}_j\otimes f^{n}_j(Q^2),
  \label{ZMVFNS}
\end{equation}
where, as in Eq.~\eqref{general}, the sum $j$ is over the gluon and the (variable) number $n\le n_f$ of quarks that are active during the evolution.  This is the so-called zero-mass variable flavour number scheme (ZM-VFNS).  However, the ZM-VFNS has the failing that it simply ignores ${\cal O}(m_h^2/Q^2)$ corrections to the coefficient functions, and hence it is inaccurate in the region where $Q^2$ is not so much greater than $m_h^2$.  Thus, strictly speaking, the ZM-VFNS does not define a scheme, but rather an approximation.

So we have two approaches where the treatment of heavy flavours is relatively simple.  The 3-flavour scheme (or FFNS), appropriate to the region $Q^2 \lesssim m_c^2$, and the ZM-VFNS, appropriate to the region $Q^2\gg m_h^2$.  Clearly, for precision parton analyses, we must use a so-called general-mass variable flavour number scheme (GM-VFNS) which smoothly connects these two well-defined regions, so as to reduce to the (3-flavour) FFNS in the low $Q^2$ limit and to the ZM-VFNS in the high $Q^2$ limit (up to possible terms of higher order in $\alpha_S$).  In particular, in Eq.~(\ref{general}) the value of $n$ increases by one each time we reach $Q^2=m_h^2$ in the evolution, and the coefficient functions $C^{n\rm GMVF}_j(Q^2/m_h^2)$ interpolate smoothly from $C^{\rm FF,3}_j(Q^2/m_h^2)$ to $C^{n_f\rm ZMVF}_j$, for the maximum value of $n=n_f$.  There is some freedom in how one does this, and in the MSTW~2008 analysis~\cite{Martin:2009iq} we use the definition of the GM-VFNS as described in Ref.~\cite{Thorne:2006qt}.\footnote{At NNLO this involves some modelling of the ${\cal O}(\alpha_S^3)$ FFNS coefficient functions which are not fully calculated.  In the GM-VFNS we only need these in the low-$Q^2$ regime and we approximate them using the known results of the small-$x$~\cite{Catani:1990eg} and threshold limits~\cite{Laenen:1998kp} (see Refs.~\cite{Alekhin:2008hc,Presti:2010pd} for recent refinements).}  We also note that actually a 5-flavour GM-VFNS is used, since we do not include the top quark in the evolution.  Note that this means that processes such as associated production of Higgs bosons with top quarks should be calculated using the 5-flavour PDFs with the full $2\to 3$ subprocess matrix elements, i.e.~$gg, q\bar q \to Ht\bar t$.  In a 6FS, one would calculate this via $t\bar{t}\to H$, introducing a top-quark PDF.  Although in principle this method would include resummed higher-order corrections $\sim [\alpha_S \ln(M_H^2/m_t^2)]^n$, we would expect that for all practical purposes\footnote{Note that for $M_H\lesssim 1$~TeV, $\alpha_S\ln(M_H^2/m_t^2)\lesssim 0.35$.} at the LHC, the 5FS approach, $gg, q\bar q \to Ht\bar t$, supplemented by NLO corrections, would be adequate, see also Section~\ref{sec:Ztot4FS}.

As just noted, the use of a GM-VFNS is nowadays essential for the determination of a set of precision PDFs from the data.\footnote{In fact we only use the GM-VFNS for structure functions in DIS, whereas hadron collider cross sections are calculated in the ZM-VFNS under the assumption that mass effects are negligible at the typically large scales.  (Low-mass fixed-target Drell--Yan production is also calculated in the ZM-VFNS, but here heavy flavours contribute $\ll 1\%$ of the total, so inaccuracies induced by using the ZM-VFNS are in practice irrelevant.)}  However, there are processes where the full mass dependence of the coefficient functions for $c$ and $b$ production processes are only known in the case where the heavy quark is treated as a final-state particle, and not as a parton in the proton.  For example, FFNS parton distributions are needed for use with the \textsc{hvqdis}~\cite{Harris:1995tu,Harris:1997zq} and \textsc{mc@nlo}~\cite{Frixione:2003ei} programs.  Thus in this paper we make available 3- and 4-flavour PDFs.  We also give an example of their use.  This is the subject of Section~\ref{sec:FFNS}.

\section{Dependence of PDFs on the heavy-quark masses~\label{sec:massdep}}
In the MSTW~2008 global parton analysis~\cite{Martin:2009iq} we used $m_c=1.40$~GeV and $m_b=4.75$~GeV, where these are the pole mass values.  Some limited justification for choosing these values was briefly given at the end of Section~3 of Ref.~\cite{Martin:2009iq}.  To summarise, there was little data constraint on $m_b$, so it was simply fixed at a value of $4.75$~GeV, close to the calculated $\overline{\rm MS}$ mass transformed to the pole mass.  The fixed value of $m_c=1.40$~GeV was close to the best-fit value at NLO if treated as a free parameter, but a little higher than the best-fit value at NNLO, and lower than the calculated $\overline{\rm MS}$ mass transformed to the pole mass.  We discuss limitations in constraints on the pole masses by transforming from the calculated $\overline{\rm MS}$ masses below in Section~\ref{sec:polemasses}.

Here, using {\it exactly} the same data sets, we repeat the global analysis for different fixed values of $m_c$ and $m_b$, but with $\alpha_S(M_Z^2)$ left as a free parameter.  The MSTW 2008 global fit~\cite{Martin:2009iq} used a wide variety of data from both fixed-target experiments and the HERA $ep$ and Tevatron $p\bar{p}$ colliders.  Neutral-current structure functions ($F_2$ and $F_L$) were included from fixed-target lepton--nucleon scattering experiments (BCDMS~\cite{Benvenuti:1989rh,Benvenuti:1989fm}, NMC~\cite{Arneodo:1996qe,Arneodo:1996kd}, E665~\cite{Adams:1996gu} and SLAC~\cite{Whitlow:1991uw,Whitlow:1990dr,Whitlow:1990gk}), low-mass Drell--Yan cross sections from the E866/NuSea experiment~\cite{Webb:2003bj,Towell:2001nh}, and charged-current structure functions ($F_2$ and $xF_3$) and dimuon cross sections from neutrino--nucleon scattering experiments (CCFR/NuTeV~\cite{Goncharov:2001qe,Tzanov:2005kr} and CHORUS~\cite{Onengut:2005kv}).  From the HERA experiments, H1 and ZEUS, data were included on neutral- and charged-current reduced cross sections ($\sigma_r^{\rm NC}$ and $\sigma_r^{\rm CC}$)~\cite{Lobodzinska:2003yd,Adloff:2000qk,Adloff:2000qj,Adloff:2003uh,Breitweg:1998dz,Chekanov:2001qu,Chekanov:2002ej,Chekanov:2003yv,Chekanov:2003vw}, the charm structure function ($F_2^c$)~\cite{Adloff:1996xq,Adloff:2001zj,Aktas:2005iw,Aktas:2004az,Breitweg:1999ad,Chekanov:2003rb,Chekanov:2007ch}, and inclusive jet production in DIS~\cite{Aktas:2007pb,Chekanov:2002be,Chekanov:2006xr}.  From the Tevatron experiments, CDF and D{\O}, Run II data were included on inclusive jet production~\cite{Abazov:2008hu,Abulencia:2007ez}, the lepton charge asymmetry from $W$ decays~\cite{Abazov:2007pm,Acosta:2005ud} and the $Z$ rapidity distribution~\cite{Abazov:2007jy,Han:2008}.  A more detailed description of the treatment of each of these data sets can be found in Ref.~\cite{Martin:2009iq}.  Note that more precise H1 and ZEUS data on $F_2^c$ (and the beauty structure function, $F_2^b$, not included originally) are now available, but we stick to fitting exactly the same data sets as in the MSTW 2008 analysis.

\subsection{Dependence on charm-quark mass $m_c$}
\begin{table}
  \centering
{\footnotesize
  \begin{minipage}{0.5\textwidth}
    (a) NLO:\\ \\
    \begin{tabular}{l|r|r|r}
      \hline\hline
      $m_c$ (GeV) & $\chi^2_{\rm global}$ & $\chi^2_{F_2^c}$ & $\alpha_S(M_Z^2)$ \\
      & (2699 pts.) & (83 pts.) & \\
      \hline
      & & & \\
      1.1  & 2729 & 263  & 0.1182 \\
      1.2  & 2625 & 188  & 0.1188 \\
      1.3  & 2563 & 134  & 0.1195 \\
 {\bf 1.4} & 2543 & 107  & 0.1202 \\
      1.45 & 2541 & 100  & 0.1205 \\
      1.5  & 2545 &  97  & 0.1209 \\
      1.6  & 2574 & 104  & 0.1216 \\
      1.7  & 2627 & 128  & 0.1223 \\
      \hline\hline
    \end{tabular}
  \end{minipage}%
  \hfill
  \begin{minipage}{0.5\textwidth}
    (b) NNLO:\\ \\
    \begin{tabular}{l|r|r|r}
      \hline\hline
      $m_c$ (GeV) & $\chi^2_{\rm global}$ & $\chi^2_{F_2^c}$ & $\alpha_S(M_Z^2)$ \\
      & (2615 pts.) & (83 pts.) & \\
      \hline
      & & & \\
      1.1  & 2498 & 113 & 0.1159 \\
      1.2  & 2463 &  88 & 0.1162 \\
      1.26 & 2456 &  82 & 0.1165 \\
      1.3  & 2458 &  82 & 0.1166 \\
 {\bf 1.4} & 2480 &  95 & 0.1171 \\
      1.5  & 2528 & 126 & 0.1175 \\
      1.6  & 2589 & 167 & 0.1180 \\
      1.7  & 2666 & 217 & 0.1184 \\
      \hline\hline
    \end{tabular}
  \end{minipage}
}
  \caption{\sf Fit quality and $\alpha_S(M_Z^2)$ for different $m_c$ values at (a) NLO and (b) NNLO.}
  \label{tab:mc}
\end{table}
The sensitivity to the charm mass of the data in the global PDF analysis, and the subset of $F_2^c$ data, in terms of the goodness-of-fit measure $\chi^2$, is shown in Table~\ref{tab:mc} at NLO and NNLO.  The global best-fit values of $m_c$, found by varying $m_c$ in steps of 0.01 GeV, are $m_c=1.45$ GeV at NLO and $m_c=1.26$ GeV at NNLO.  Recall that in Section~3 of Ref.~\cite{Martin:2009iq} we instead quoted best-fit values of $m_c=1.39$~GeV at NLO and $m_c=1.27$~GeV at NNLO.  In fact, the more detailed analysis performed here, by scanning $m_c$ in steps of 0.01 GeV, revealed a rather flat minimum of the global $\chi^2$ versus $m_c$, with a slight double-minimum structure at NLO (the two minima being at 1.39~GeV and 1.45~GeV).  Hence the best-fit values we quote here, of $m_c=1.45$~GeV at NLO and $m_c=1.26$~GeV at NNLO, differ slightly from the best-fit values given in Ref.~\cite{Martin:2009iq}.  Note from Table~\ref{tab:mc}(a) that there is a clear correlation between $m_c$ and the value of $\alpha_S(M_Z^2)$ obtained from the NLO fit, and that from Table~\ref{tab:mc}(b) this correlation is reduced at NNLO.

\begin{figure}
  \centering
  \begin{minipage}{0.8\textwidth}
    (a)\\
    \includegraphics[width=\textwidth]{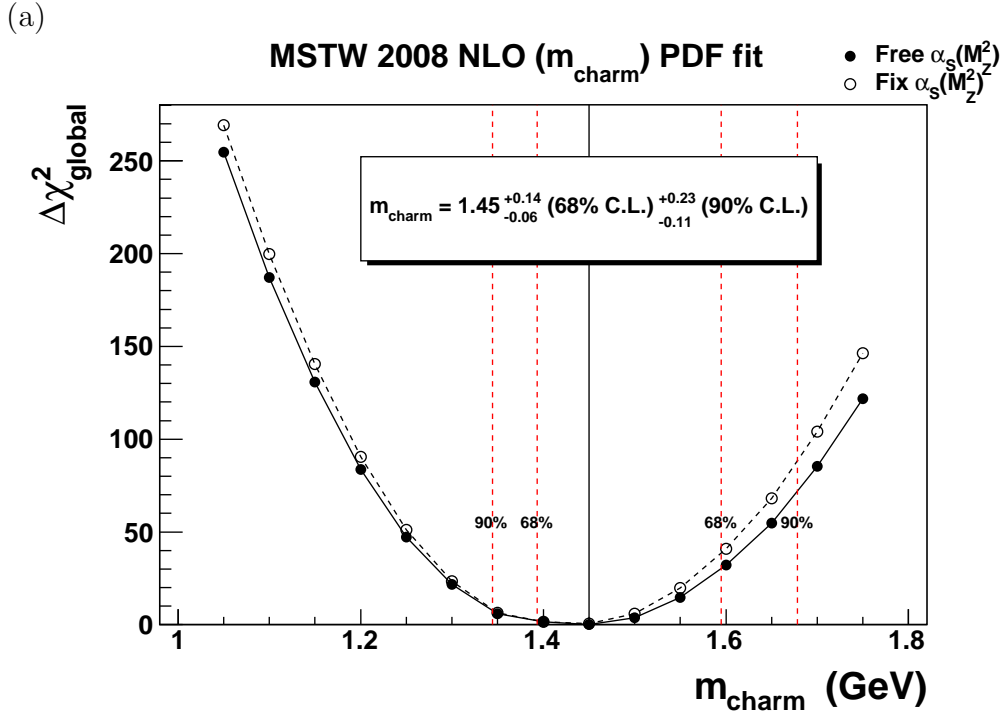}
  \end{minipage}
  \begin{minipage}{0.8\textwidth}
    (b)\\
    \includegraphics[width=\textwidth]{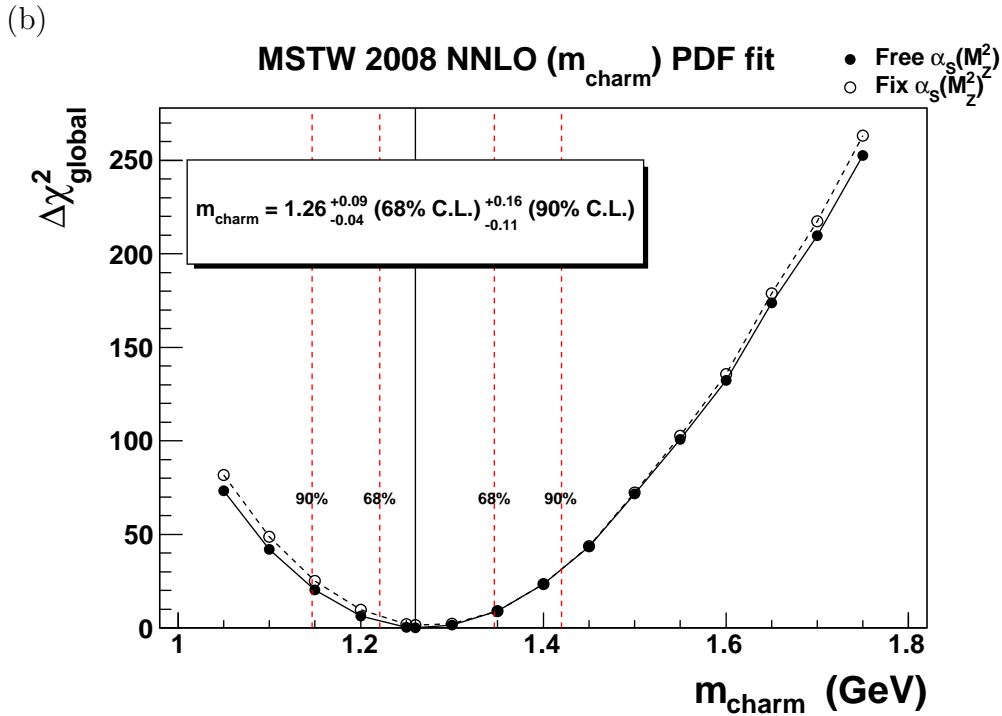}
  \end{minipage}
  \caption{\sf Dependence on $m_c$ of the global $\chi^2$, relative to the value at the best-fit $m_c$, at (a)~NLO and (b)~NNLO, fitting to the same data sets used in the MSTW~2008 global analysis~\cite{Martin:2009iq}.  The values quoted for $m_c$ correspond to treating $\alpha_S(M_Z^2)$ as a free parameter, and the corresponding global $\chi^2$ values are shown by the \emph{closed points} ($\bullet$) joined by the \emph{continuous curves}.  The $68\%$ ($1\sigma$) and $90\%$ confidence-level (C.L.) uncertainties on $m_c$ are indicated by the \emph{vertical dashed lines}.  The \emph{open points} ($\circ$), joined by the \emph{dashed curves}, are the global $\chi^2$ values with $\alpha_S(M_Z^2)$ held fixed.}
  \label{fig:globalmc}
\end{figure}
In Fig.~\ref{fig:globalmc} we show the NLO and NNLO global $\chi^2$ profiles as a function of $m_c$.  We see that at NLO there is a preference for $m_c=1.45$~GeV, with a 1$\sigma$ uncertainty of $^{+0.14}_{-0.06}$~GeV.  The global data prefer a lower value of $m_c$ at NNLO, namely $m_c=1.26$~GeV, with a 1$\sigma$ uncertainty of $^{+0.09}_{-0.04}$~GeV.  (For the higher values of $m_c$, the NNLO fit considerably undershoots the moderate $Q^2$ deep-inelastic data.)  Here, the 1$\sigma$ uncertainty is obtained in exactly the same way as described for the uncertainty on $\alpha_S(M_Z^2)$ in Ref.~\cite{Martin:2009bu}, i.e.~by examining the $\chi^2$ profile of each data set included in the global fit.
\begin{figure}
  \centering
  \begin{minipage}{0.8\textwidth}
    (a)\\
    \includegraphics[width=\textwidth]{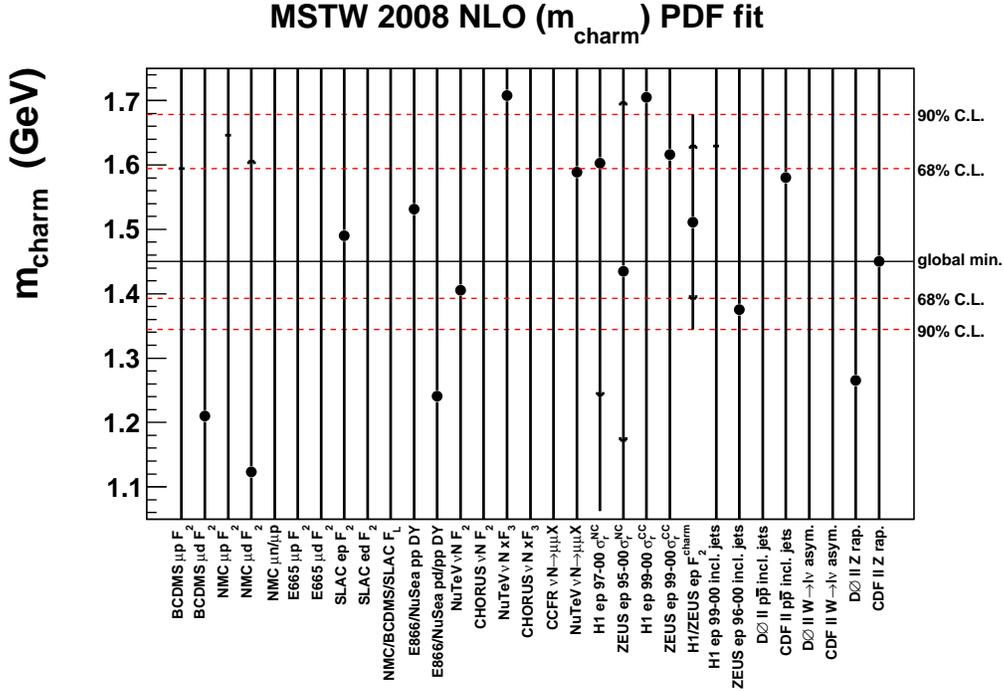}
  \end{minipage}
  \begin{minipage}{0.8\textwidth}
    (b)\\
    \includegraphics[width=\textwidth]{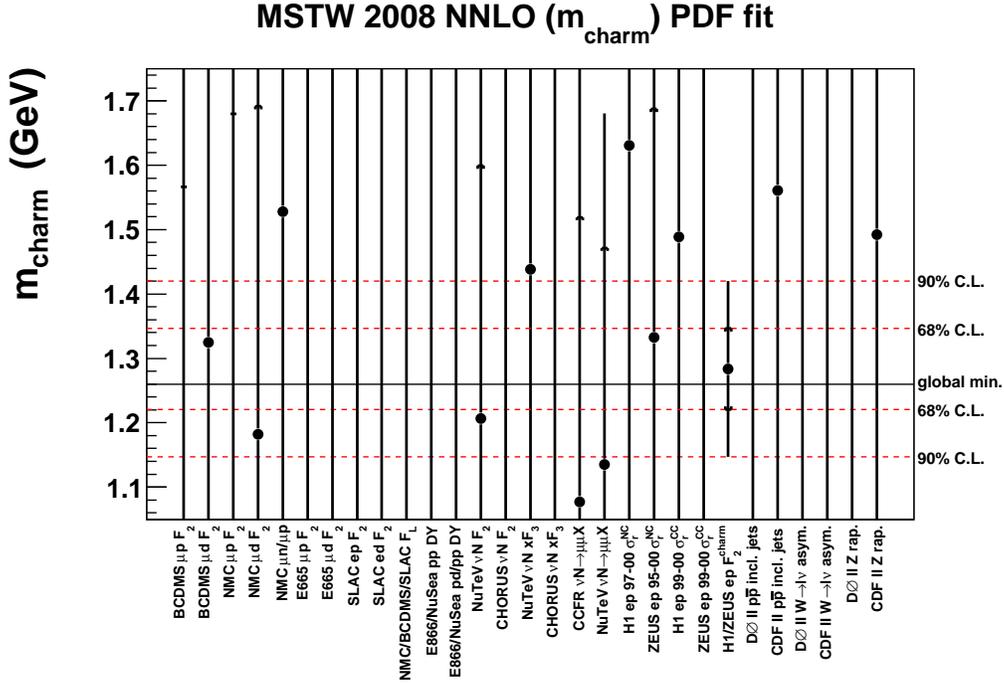}
  \end{minipage}
  \caption{\sf Ranges of $m_c$ for which data sets are described within their $90\%$ C.L.~limit (\emph{outer error bars}) or $68\%$ C.L.~limit (\emph{inner error bars}) in the (a)~NLO and (b)~NNLO global fits, with $\alpha_S(M_Z^2)$ as a free parameter.  The \emph{points} ($\bullet$) indicate the values of $m_c$ favoured by each individual data set, that is, the values for which the $\chi^2$ is minimised.  The experimental uncertainty on $m_c$, indicated by the \emph{horizontal dashed lines}, is chosen to ensure that all data sets are described within their $68\%$ or $90\%$ C.L.~limits defined by Eq.~(2) of Ref.~\cite{Martin:2009bu}.}
  \label{fig:rangemc}
\end{figure}
The distinguishing power of the various data sets is shown in Fig.~\ref{fig:rangemc}.  Here, the points ($\bullet$) indicate the values of $m_c$ for which the $\chi^2$ for each data set is minimised (within the context of the global fit), while the inner error bars extend across the 68\% confidence-level (C.L.)~region and the outer error bars extend across the 90\% C.L.~region (see Section 4 of Ref.~\cite{Martin:2009bu} for the precise definitions of the 68\% and 90\% C.L.~limits).  As may be expected, the $F_2^c(x,Q^2)$ data of the H1 and ZEUS collaborations are the most discriminating.
\begin{figure} 
  \centering
  \begin{minipage}{0.5\textwidth}
    (a) MSTW 2008 NLO:\\
    \includegraphics[width=\textwidth]{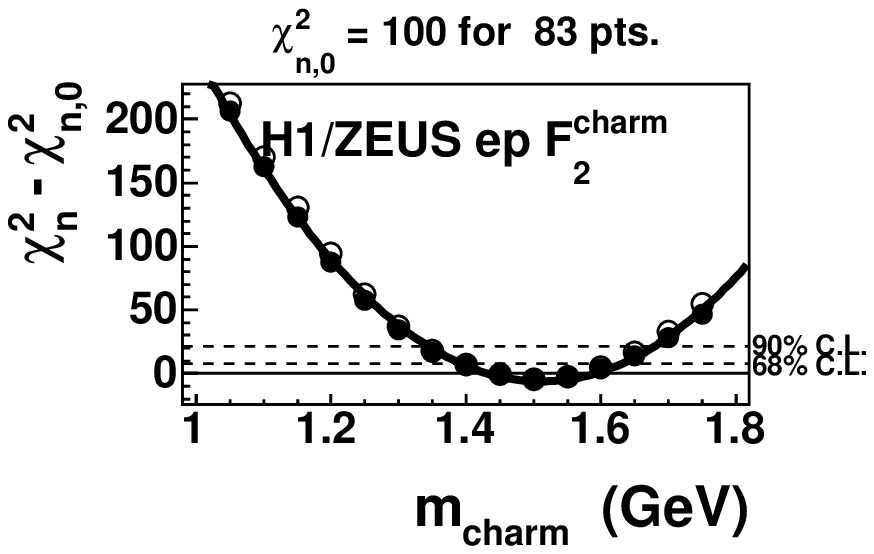}
  \end{minipage}%
  \begin{minipage}{0.5\textwidth}
    (b) MSTW 2008 NNLO:\\
    \includegraphics[width=\textwidth]{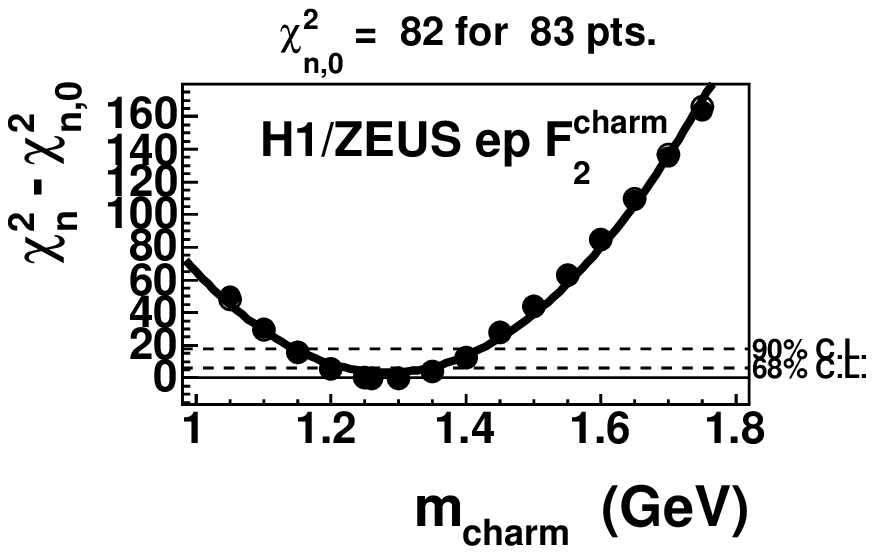}
  \end{minipage}
  \caption{\sf Dependence on $m_c$ of the $\chi^2$ for HERA data~\cite{Adloff:1996xq,Adloff:2001zj,Aktas:2005iw,Aktas:2004az,Breitweg:1999ad,Chekanov:2003rb,Chekanov:2007ch} on the charm structure function, $F_2^c$, relative to the value at the global best-fit $m_c$, at (a)~NLO and (b)~NNLO.  The \emph{closed points} ($\bullet$) have $\alpha_S(M_Z^2)$ as a free parameter, and are fitted to a quadratic function of $m_c$ shown by the continuous curves, while the \emph{open points} ($\circ$, mostly hidden) have $\alpha_S(M_Z^2)$ held fixed at the MSTW 2008 value.  The \emph{horizontal dashed lines} in the plots indicate the $68\%$ and $90\%$ C.L.~limits, determined according to a ``hypothesis-testing'' criterion, see Section 4 of Ref.~\cite{Martin:2009bu}.}
  \label{fig:chi2charm}
\end{figure}
Indeed, they dominate the determination of $m_c$, as can be seen from Fig.~\ref{fig:chi2charm}, which shows $\chi^2$ versus $m_c$ for the $F_2^c$ data alone.  For the HERA inclusive data, changes in the value of $m_c$  are partially compensated by changes in the gluon and in $\alpha_S$.  The NMC data prefer a lower value of $m_c$ giving a quicker evolution near threshold.  Similarly the BCDMS data mainly prefer a lower $m_c$, but only due to the correlation with a lower value of $\alpha_S$.  Supplementary plots of the $\chi^2$ profiles versus $m_c$ for all data sets in the global fit are available from Ref.~\cite{mstwpdf}.  The experimental uncertainty on $m_c$, given in Fig.~\ref{fig:globalmc} and indicated by the horizontal dashed lines in Fig.~\ref{fig:rangemc}, is chosen to ensure that all data sets are described within their 68\% or 90\% C.L.~limits.

\begin{figure}
  \centering
  \includegraphics[width=0.5\textwidth]{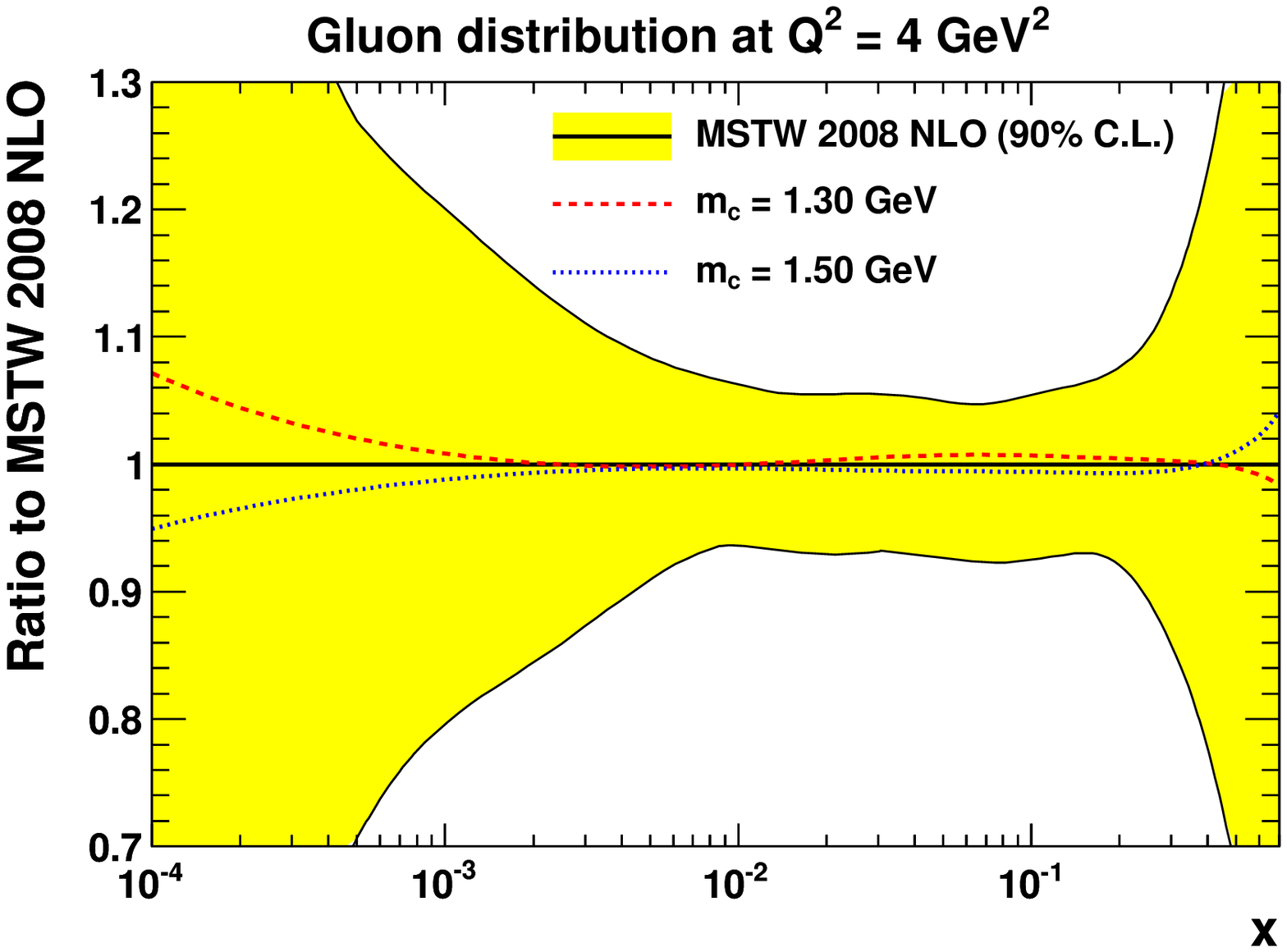}%
  \includegraphics[width=0.5\textwidth]{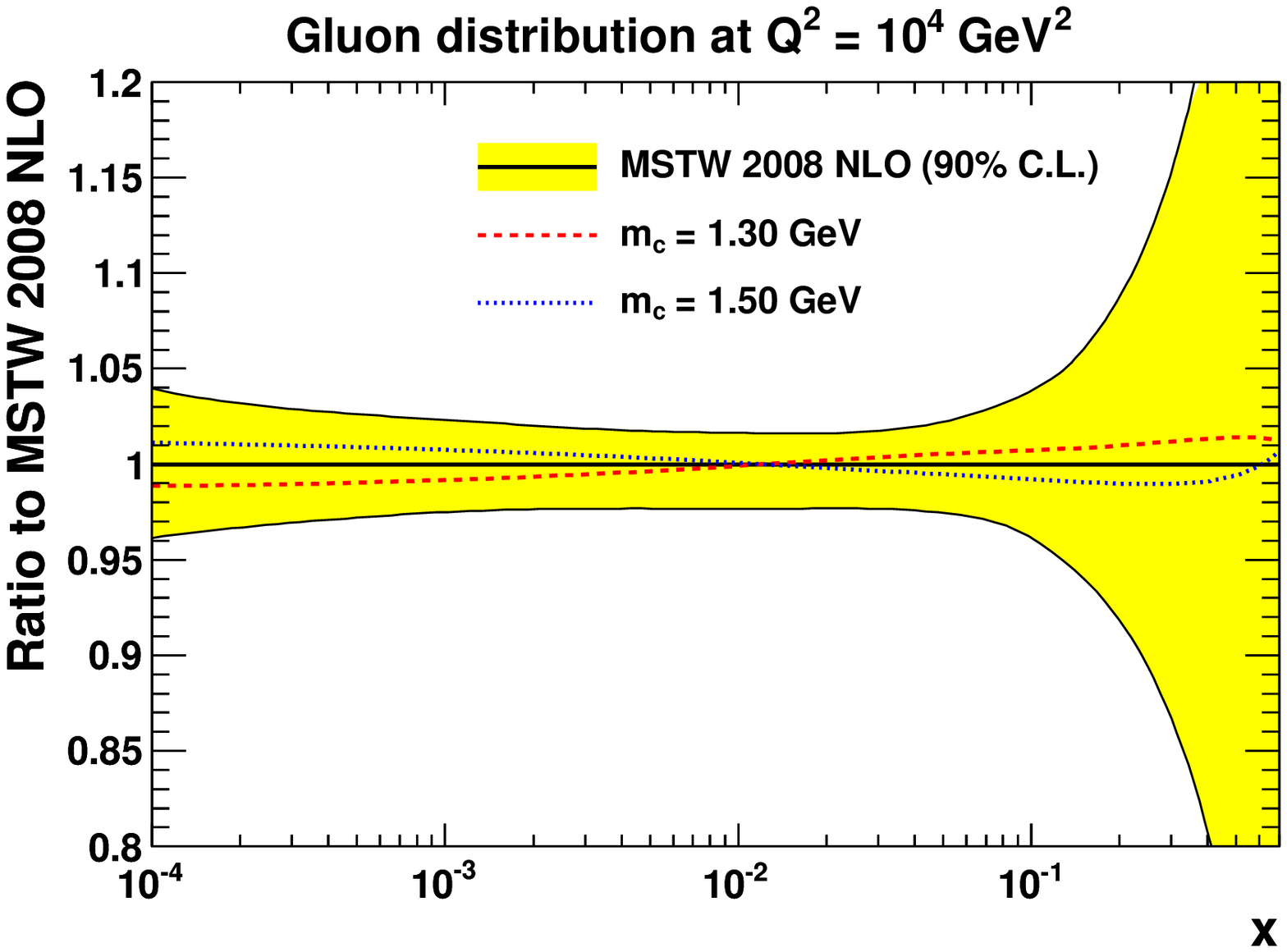}\\
  \includegraphics[width=0.5\textwidth]{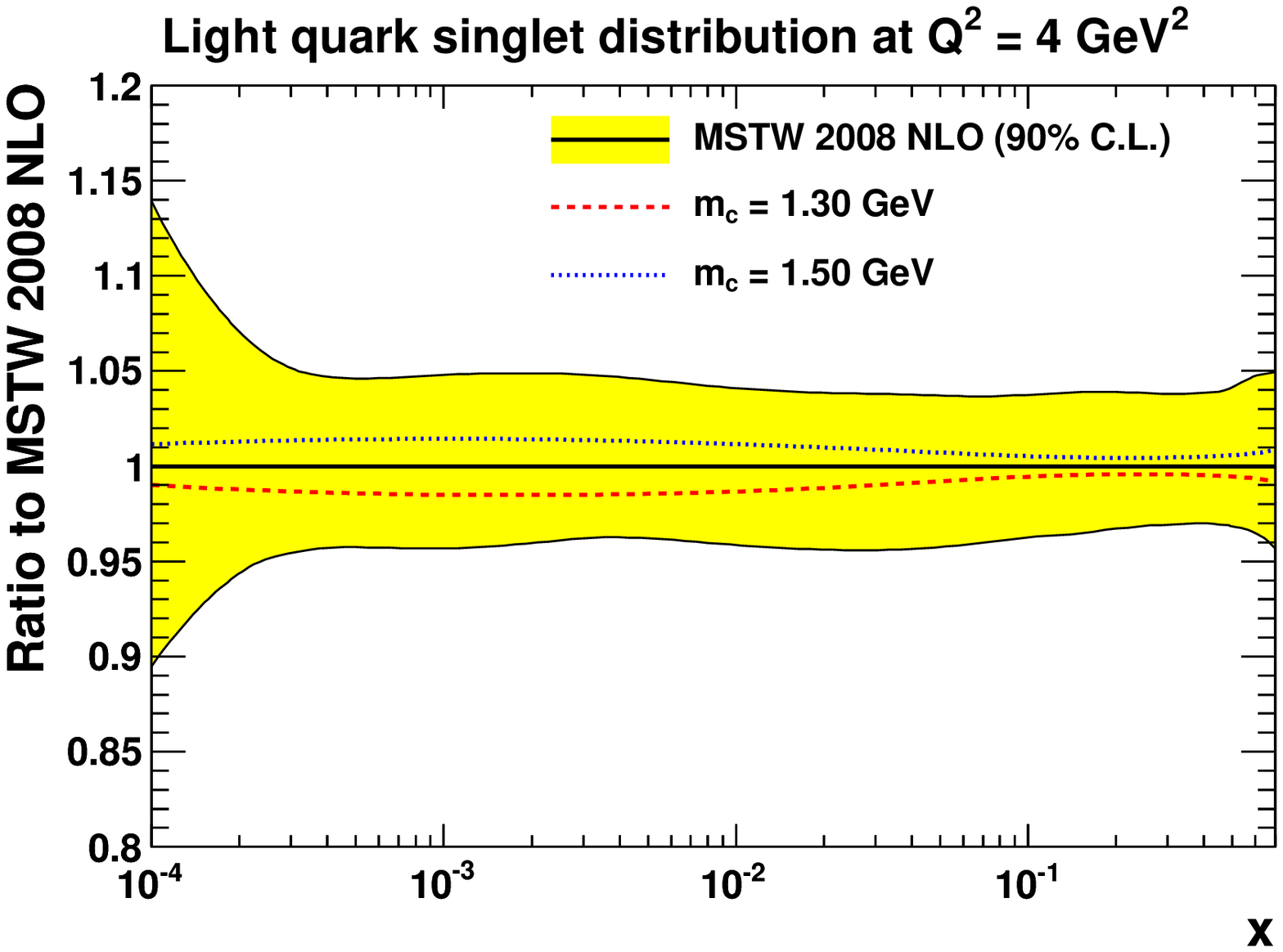}%
  \includegraphics[width=0.5\textwidth]{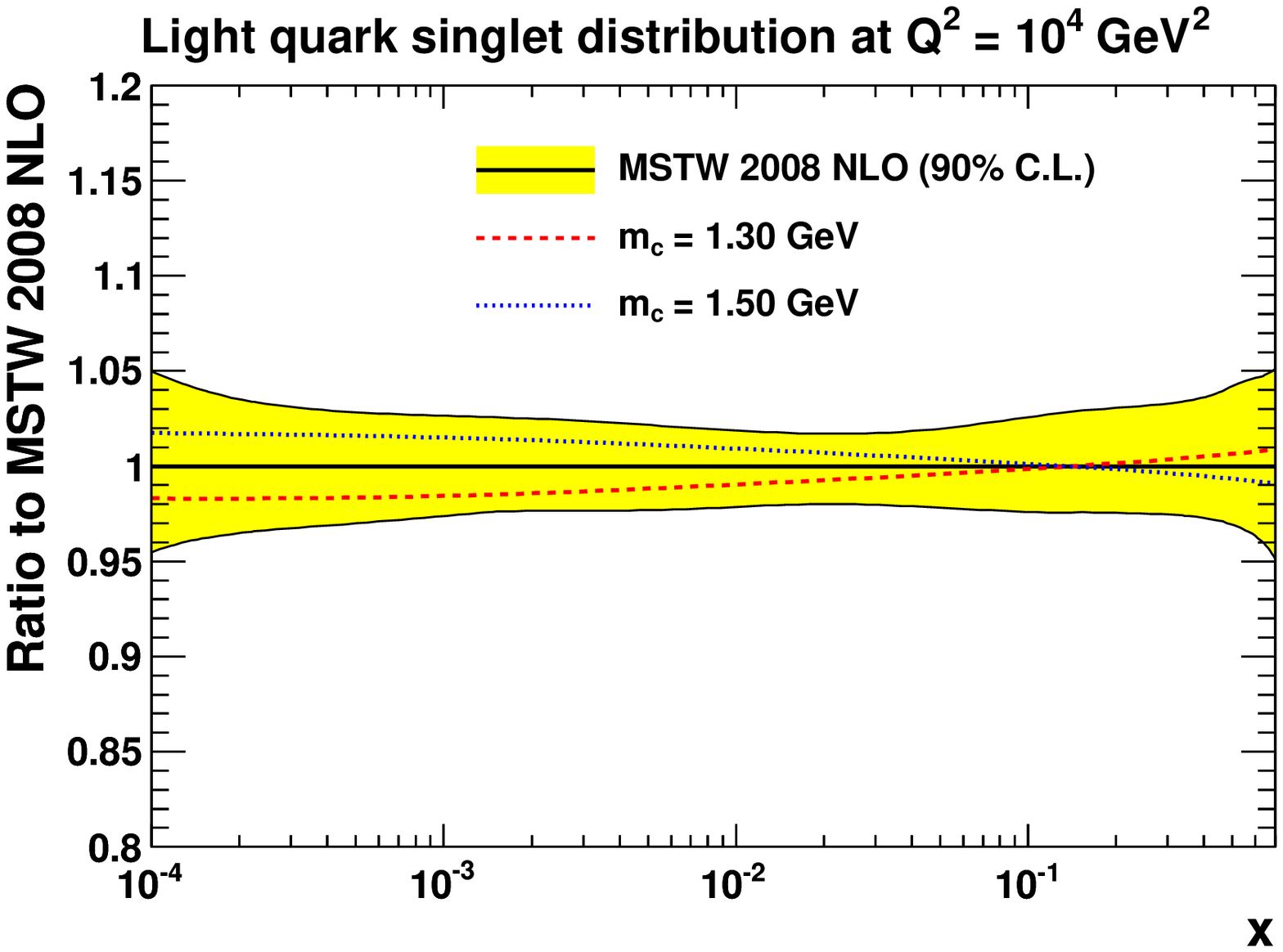}\\
  \includegraphics[width=0.5\textwidth]{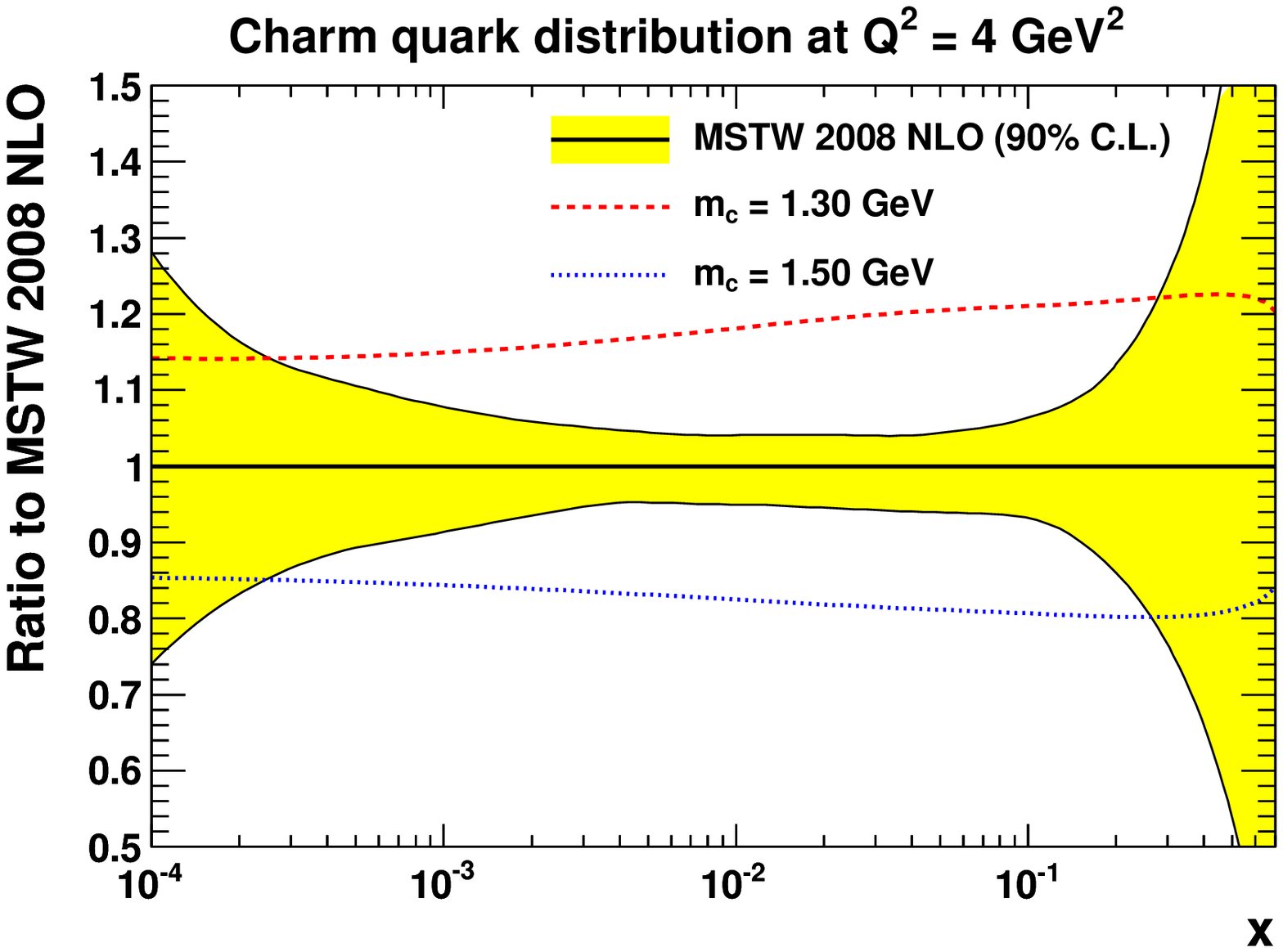}%
  \includegraphics[width=0.5\textwidth]{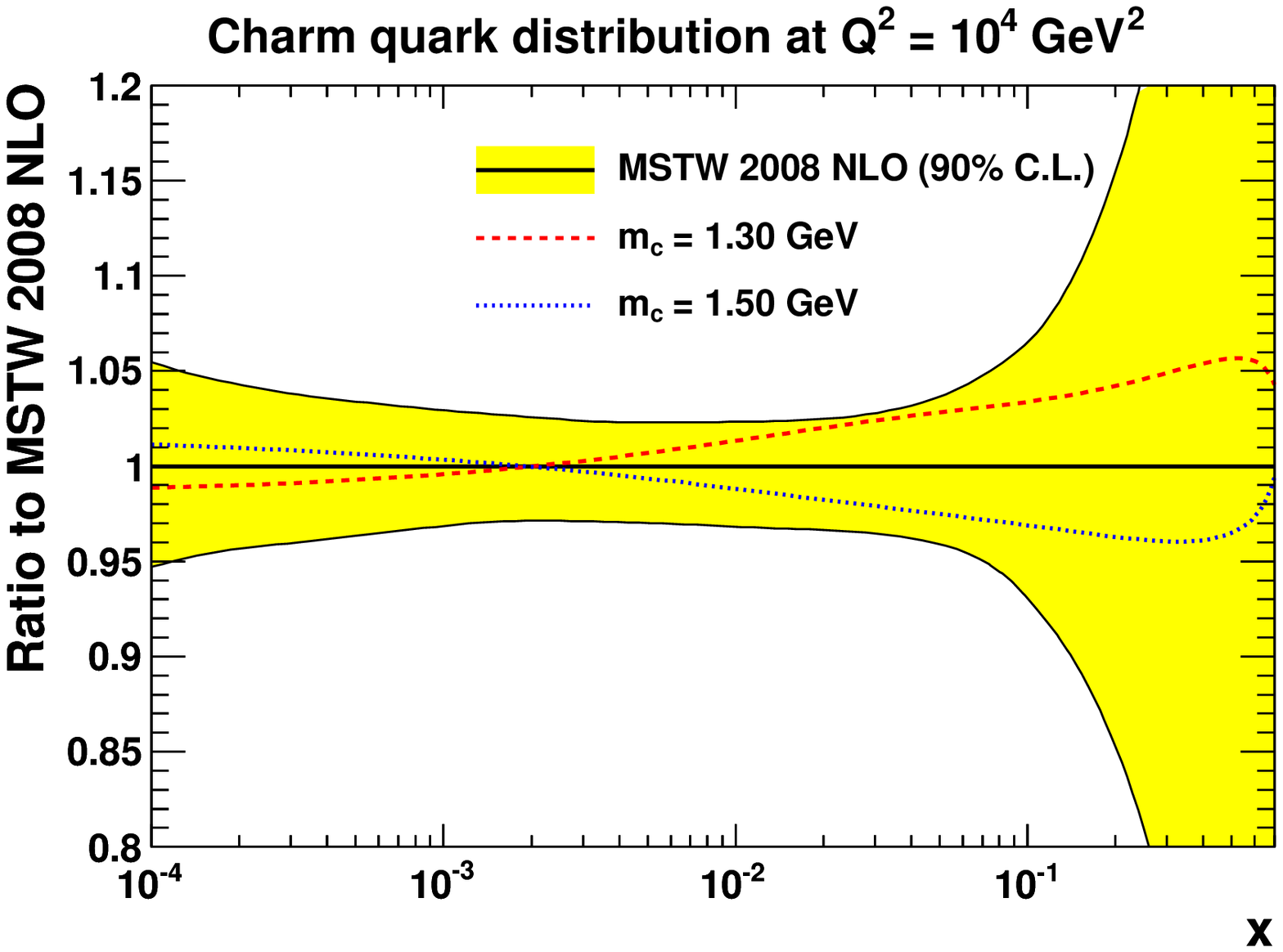}
  \caption{\sf $m_c$ dependence of gluon, singlet and charm distributions at NLO at two different $Q^2$ values, $4$~GeV$^2$~(\emph{left}) and $10^4$~GeV$^2$~(\emph{right}), compared to the $90\%$ C.L.~PDF uncertainties.}
  \label{fig:mcdep}
\end{figure}
The dependence of the PDFs on the value taken for $m_c$ is shown in Fig.~\ref{fig:mcdep}.  The plots show the ratio of the NLO PDFs obtained with $m_c=1.3$~GeV and $m_c=1.5$~GeV to those of the MSTW~2008 analysis (which assumed $m_c=1.4$ GeV) at scales of $Q^2=4$~GeV$^2$ and $Q^2=10^4$~GeV$^2$.  We see that the ratios lie well within the 90\% C.L.~PDF uncertainty.  The exception is the charm distribution at $Q^2=4$~GeV$^2$ where, as expected, the different charm transition points have an appreciable effect.

\subsection{Dependence on bottom-quark mass $m_b$}
\begin{table}
  \centering
{\footnotesize
  \begin{tabular}{l||r|r|r|r||r|r|r|r}
    \hline\hline
    & \multicolumn{4}{c||}{NLO} & \multicolumn{4}{c}{NNLO} \\
    \hline
    $m_b$ (GeV) & $\chi^2_{\rm global}$ & $\chi^2_{F_2^b}$ & $\chi^2_{{\rm global}+F_2^b}$ & $\alpha_S(M_Z^2)$ & $\chi^2_{\rm global}$ & $\chi^2_{F_2^b}$ & $\chi^2_{{\rm global}+F_2^b}$ & $\alpha_S(M_Z^2)$ \\
    & (2699 pts.) & (12 pts.) & (2711 pts.) & & (2615 pts.) & (12 pts.) & (2627 pts.) & \\
    \hline
    & & & & & & & & \\
    4.00 & 2537 & 20 & 2557 & 0.1202 & 2477 & 21 & 2498 & 0.1171 \\
    4.25 & 2539 & 13 & 2552 & 0.1202 & 2478 & 15 & 2493 & 0.1171 \\
    4.50 & 2541 & 8.9 & 2550 & 0.1202 & 2478 & 11 & 2489 & 0.1171 \\
    {\bf 4.75} & 2543 & 7.4 & 2550 & 0.1202 & 2480 & 8.8 & 2489 & 0.1171 \\
    5.00 & 2545 & 7.6 & 2553 & 0.1202 & 2481 & 6.9 & 2488 & 0.1170 \\
    5.25 & 2547 & 7.6 & 2555 & 0.1201 & 2483 & 7.7 & 2491 & 0.1170 \\
    5.50 & 2549 & 8.0 & 2557 & 0.1201 & 2485 & 7.9 & 2493 & 0.1170 \\
    \hline\hline
  \end{tabular}
}
  \caption{\sf Fit quality and $\alpha_S(M_Z^2)$ for different $m_b$ values at NLO and NNLO.  Note that the $F_2^b$ data~\cite{Aaron:2009ut} are not included in the global fit.}
  \label{tab:mb}
\end{table}
We repeated the whole exercise for $m_b$. That is, we performed a series of global fits for different values of $m_b$. The results are shown in Table~\ref{tab:mb}.  Now there is less sensitivity to the value of the quark mass, and essentially no correlation between $m_b$ and $\alpha_S(M_Z^2)$.  Indeed, the global $\chi^2$ stays fairly flat all the way down to $m_b=3$~GeV.  For the lower values of $m_b$, there is a slightly better description of the HERA data, including $F_2^c(x,Q^2)$.  A similar conclusion holds at NNLO, but with about half the change in global $\chi^2$, that is, even less sensitivity to $m_b$.

\begin{figure}
  \begin{center}
    \includegraphics[width=\textwidth]{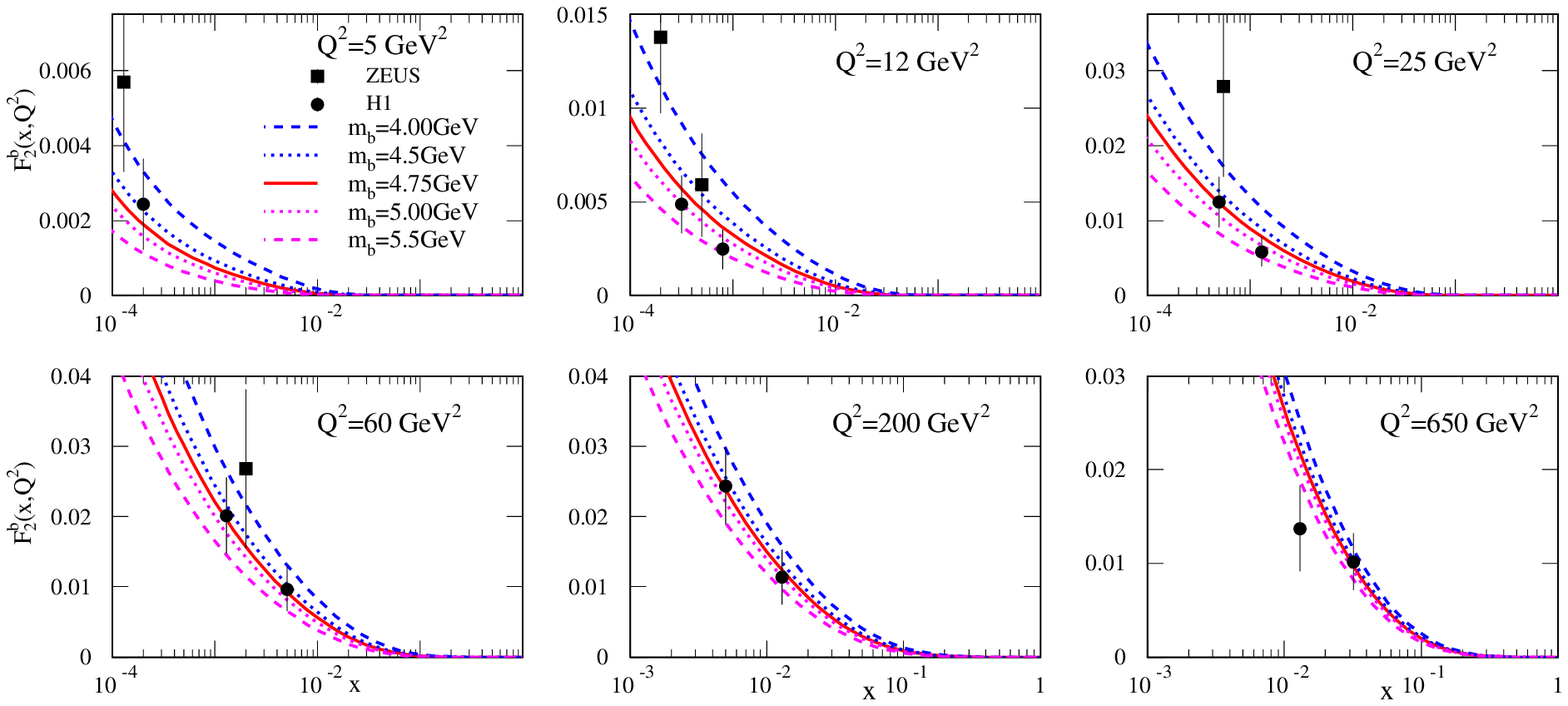}
    \caption{\sf Comparison of the HERA $F_2^b(x,Q^2)$ data~\cite{Aaron:2009ut,Abramowicz:2010zq} with NNLO predictions obtained using PDFs from global fits with different values of $m_b$.  Note that the $F_2^b$ data are not included in these global fits.  The \emph{curves}, in decreasing order, correspond to $m_b=4,~4.5, ~4.75, ~5, ~5.5$ GeV.}
    \label{fig:F2b}
  \end{center}
\end{figure}
Note that the HERA $F_2^b(x,Q^2)$ data~\cite{Aktas:2004az,Aktas:2005iw,Aaron:2009ut,Chekanov:2009kj,Abramowicz:2010zq} are not included in the global fits, neither in MSTW~2008~\cite{Martin:2009iq}, nor in the fits described above.  In Fig.~\ref{fig:F2b} we compare the predictions of the NNLO fits with varying $m_b$ values to the $F_2^b$ data from H1~\cite{Aaron:2009ut}.  We also show a few of the (less precise) ZEUS data points~\cite{Abramowicz:2010zq}.  In Table~\ref{tab:mb} we give the $\chi^2$ for the H1 $F_2^b$ data accounting\footnote{We use Eqs.~(38--40) of Ref.~\cite{Martin:2009iq}, noting that there is a typo in Eq.~(40) of that paper and $\sigma_{n,i}^{\rm uncorr.}$ should appear squared in the expression for $A_{kk^\prime}$.} for all 24 sources of correlated systematic uncertainty, and we also give the simple addition with the global $\chi^2$.  We see that the $F_2^b$ data show a slight preference for $m_b\approx 4.75$~GeV at NLO and $m_b\approx 5.00$~GeV at NNLO.  The global fits (including the $F_2^b$ data, obtained by simply adding the $\chi^2$ values), would prefer $m_b\approx 4.50$--4.75~GeV at NLO and $m_b\approx 4.50$--5.00~GeV at NNLO, although admittedly the $\chi^2$ profiles are quite flat, see Table~\ref{tab:mb}.  We conclude that the global data do not meaningfully constrain $m_b$, and that, in view of Table~\ref{tab:mb} and Fig.~\ref{fig:F2b}, the choice $m_b=4.75$~GeV made in the MSTW~2008~\cite{Martin:2009iq} analysis is completely satisfactory.

\begin{figure}
  \centering
  \begin{minipage}{0.5\textwidth}
    (a)\\
    \includegraphics[width=\textwidth]{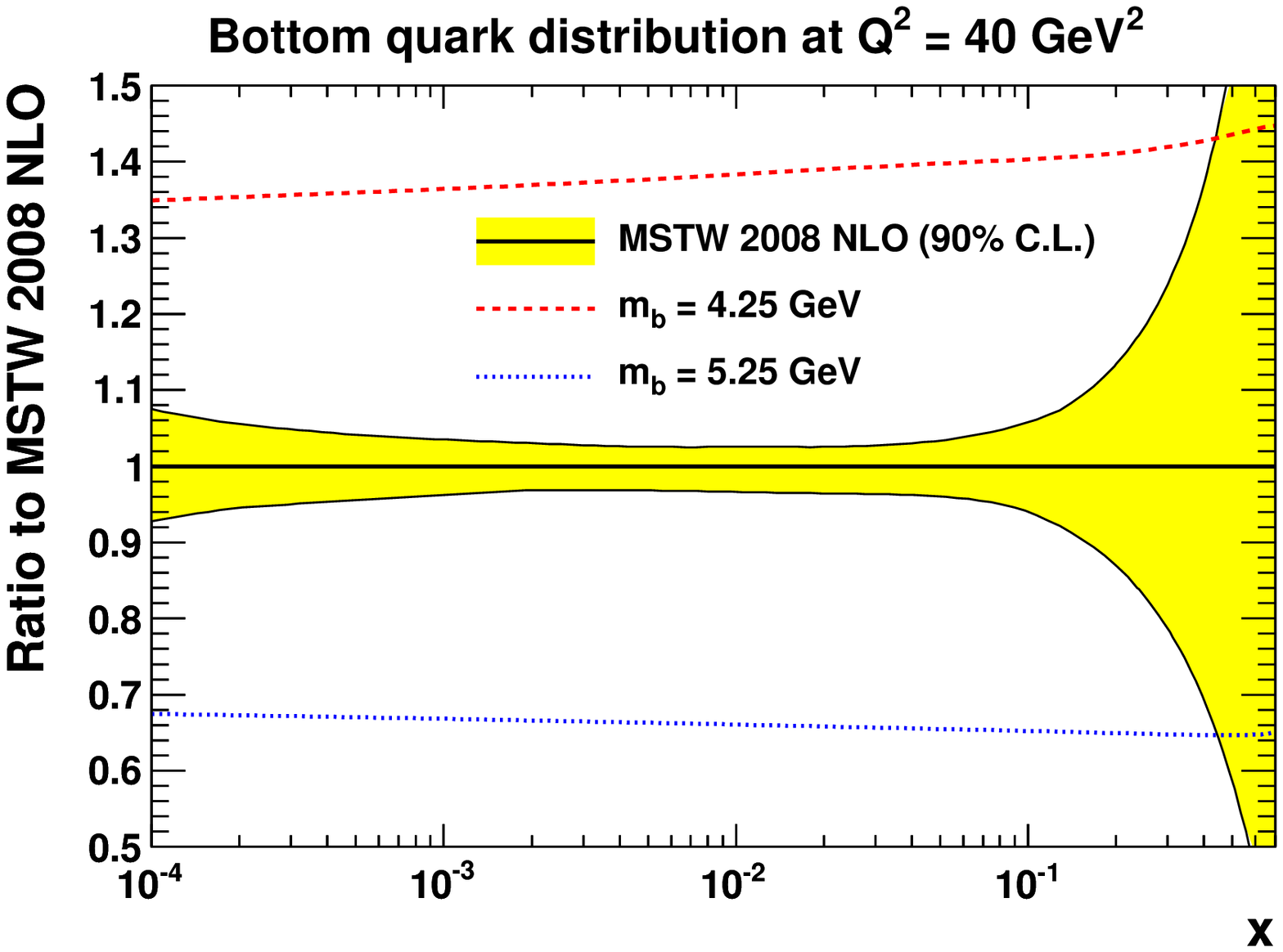}
  \end{minipage}%
  \begin{minipage}{0.5\textwidth}
    (b)\\
    \includegraphics[width=\textwidth]{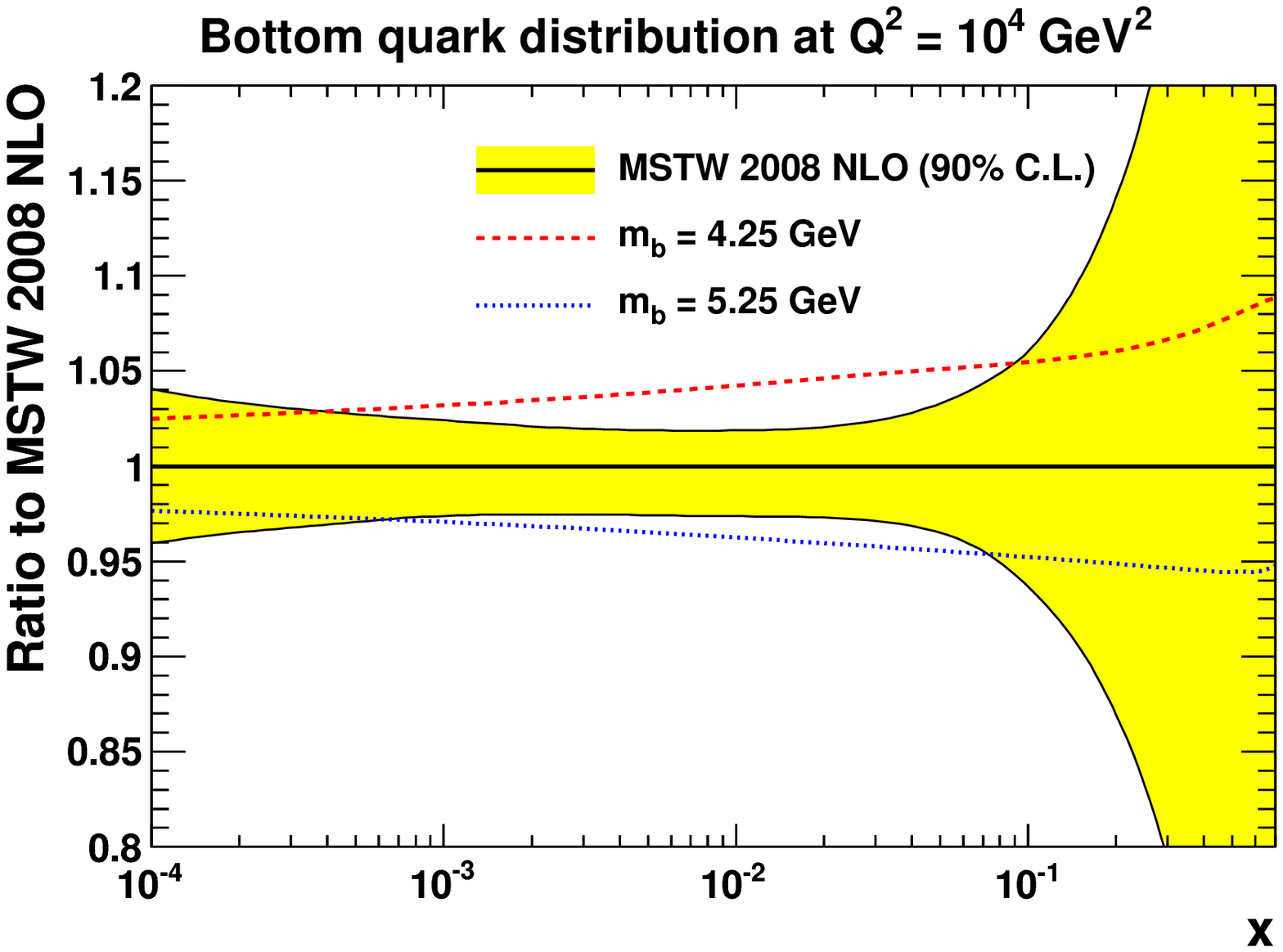}
  \end{minipage}
  \caption{\sf $m_b$ dependence of bottom-quark distribution at NLO at two different $Q^2$ values, (a)~$40$~GeV$^2$ and (b)~$10^4$~GeV$^2$, compared to the $90\%$ C.L.~PDF uncertainties.}
  \label{fig:mbdep}
\end{figure}
The dependence of the bottom-quark distribution on the value taken for $m_b$ is shown in Fig.~\ref{fig:mbdep}.  The plots show the ratio of the NLO PDFs obtained with $m_b=4.25$~GeV and $m_b=5.25$~GeV to those of the MSTW~2008 analysis (which assumed $m_b=4.75$ GeV) at scales of $Q^2=40$~GeV$^2$ and $Q^2=10^4$~GeV$^2$.  We see that the $b$-quark ratios lie well outside the 90\%~C.L.~PDF uncertainty at $Q^2=40$~GeV$^2$ and even at $Q^2=10^4$~GeV$^2$.  The gluon distribution and the other quark flavours are hardly affected by the value of $m_b$ (not shown here).

\subsection{Pole masses of heavy quarks} \label{sec:polemasses}
Note that in our analyses $m_c$ refers to the {\it pole} mass of the charm quark.  Both the transition matrix elements defining the boundary conditions for heavy-quark evolution~\cite{Buza:1996wv} and the heavy-quark coefficient functions used to define the GM-VFNS~\cite{Buza:1996wv} are defined using the on-shell renormalisation scheme for the heavy quark.  Unlike $\alpha_S(\mu^2)$, the pole mass $m_c$ is, in principle, a physically-defined (spoilt by confinement which gives power corrections, see below) quantity which is stable to the order of perturbation theory.  So the values that we obtain in the NLO and NNLO analyses should, in principle, be the same.  However, we see from Fig.~\ref{fig:globalmc} that the values are only marginally compatible.  This is due to a variety of reasons.  The perturbative expansion of heavy-flavour coefficient functions is not as convergent as one might hope, with large corrections near threshold and in the small-$x$ limit.  So even though the pole mass might, in principle, be the same quantity at two different perturbative orders, the calculation at lower orders is missing significant corrections at higher orders which will affect the value of the pole mass obtained in an extraction from data.  Perhaps most important in this respect is the large negative boundary condition (at $Q^2=m_h^2$) for the heavy-flavour distribution at NNLO, which has the effect of flattening the slope $\partial F_2^h/\partial\ln Q^2$ (see Ref.~\cite{Thorne:2006qt} for details) and implies a lower mass to correct for this.  There is also more potential variation in the GM-VFNS definition at NLO than NNLO, and there would be slightly more stability in preferred pole mass in, for example, the ``optimal'' scheme introduced in Ref.~\cite{Thorne:2010pa} (although this does not remove the variation and is the subject for a future study). Finally, as briefly noted above and discussed further below, the definition of the pole mass is contaminated by non-perturbative corrections.  These are reflected in the divergence of the perturbation theory and power corrections are, in practice, difficult to disentangle from higher-order corrections, so extractions at different orders will have a variation from this source.  We now examine this issue in more detail.

Let us consider the values we extract and use as default.  In the \emph{Review of Particle Physics}~\cite{Amsler:2008zzb}, the authors quote values for the ``{\it running}'' masses, $m_c(\mu=m_c)$ and $m_b(m_b)$, in the $\overline{\rm MS}$ scheme:
\begin{equation}
  m_c(\mu=m_c)~=~1.27^{~+0.07}_{~-0.11}~~{\rm GeV},\qquad m_b(m_b)~=~4.20~^{+0.17}_{-0.07}~~{\rm GeV},
  \label{eq:running}
\end{equation}
which they evaluate from a whole series of independent determinations.  This implies a precise determination of masses.\footnote{There are numerous individual determinations of both charm- and bottom-quark masses in the $\overline{\rm MS}$ scheme which are actually even more precise than the above---for recent examples see~Refs.~\cite{Chetyrkin:2009fv,Narison:2010cg}.  However, these can vary from each other by considerably more than the quoted uncertainties, so in the \emph{Review of Particle Physics}~\cite{Amsler:2008zzb} account is taken of the spread, as well as the individual uncertainties.}  The best precision of the pole mass should be obtained by using this $\overline{\rm MS}$ value (the best determinations coming from NNNLO calculations) with the most accurate conversion factor, also at NNNLO.  However, the conversion of the ``{\it running}'' mass to the pole mass is problematic, since it relies on a very weakly convergent perturbative expansion.  Using our NNLO value of $\alpha_S$, the term-by-term conversion factor (obtained from the formula in Ref.~\cite{Amsler:2008zzb}), starting at zeroth order and going to the highest known order of $\alpha_S^3$~\cite{Chetyrkin:1999qi,Melnikov:2000qh}, for the bottom-quark mass is $(1+0.095+0.045+0.036)$ and for charm is $(1+0.16+0.14+0.18)$.\footnote{Use of an NNNLO value of $\alpha_S$ would make little difference.  As shown in Ref.~\cite{Kuhn:2008zzd}, NNNLO extractions of $\alpha_S(M_Z^2)$ are very similar to the corresponding NNLO values, and the slightly quicker running at NNNLO would marginally increase $\alpha_S$ at the scale of the charm- and bottom-quark masses, i.e.~the convergence of the perturbative series would likely be very slightly worse.}  The former implies that convergence is ceasing, while the latter has no real convergence at all, since it relies on a much higher value of $\alpha_S$ (and larger coefficients due to less cancellation between individual terms as the number of light quark flavours decreases~\cite{Chetyrkin:1999qi,Melnikov:2000qh}).

Indeed, the contamination by renormalons of the perturbative series for the transformation of a running quark mass in the $\overline{\rm MS}$ scheme to the pole mass has been known for a long time~\cite{Bigi:1994em,Beneke:1994sw}.  These papers estimate the intrinsic ambiguity of the perturbative series due to infrared contributions, translating into an uncertainty on the pole mass of at least $0.05~{\rm GeV}$, but more likely of order 0.1--0.2~GeV.  The best estimate one can obtain from the perturbative series is usually defined by including all terms in the series until they cease to fall in magnitude from one order to the next.  For the bottom quark, under the plausible assumption that the unknown  ${\cal O}(\alpha_S^4)$ term is similar in size to the calculated ${\cal O}(\alpha_S^3)$ term, this would give a pole mass of $m_b = 4.9$~GeV.  The uncertainty on this value from the conversion is either the estimate from Ref.~\cite{Beneke:1994sw} or roughly the size of the last term included in the series, i.e.~$0.15$~GeV.  The good correspondence of the two estimates of the uncertainty is expected, and suggests that the series is indeed truncated at the correct point.  If this uncertainty is added (approximately) in quadrature with the uncertainty on the $\overline{\rm MS}$ scheme mass, Eq.~\eqref{eq:running}, we obtain $m_b = 4.9\pm0.2$~GeV.

The determination of the charm-quark pole mass by conversion from the $\overline{\rm MS}$ scheme mass is more problematic since, as noted above, the perturbative series displays no clear convergence at all.  However, as discussed in Ref.~\cite{Hoang:2005zw}, the leading renormalon contribution is the same for different masses, i.e.~$\delta m^2 \propto \Lambda_{\rm QCD}^2$, independent of quark flavour.  This means the difference between $m_b$ and $m_c$ can be well determined, and turns out to be $3.4$~GeV with a very small uncertainty~\cite{Hoang:2005zw}.  Using this result we obtain a best determination of the charm-quark pole mass of $m_c=1.5\pm 0.2$~GeV.  There is an uncertainty on $m_b-m_c$ due to the fact that the non-dominant renormalon contributions, $\delta m^2 \propto \Lambda_{\rm QCD}^4/m^2$ and beyond, do not cancel.  However, the uncertainty on $m_b-m_c$ is small compared to the $0.2$~GeV uncertainty on $m_b$ and hence does not affect the uncertainty on $m_c$ at the quoted accuracy when added in quadrature.

These pole mass values obtained from $\overline{\rm MS}$ conversion:
\begin{equation}
  m_c~=~1.5~\pm0.2~~{\rm GeV},\qquad m_b~=~4.9~\pm0.2~~{\rm GeV},
  \label{eq:polemasses}
\end{equation}
are slightly high compared with our default values ($1.4$~GeV and $4.75$~GeV) and our best-fit values of $m_c$, particularly at NNLO, but even in the most discrepant case the error bars overlap.  Combining our best-fit $m_c$ values and the pole mass determination given in Eq.~\eqref{eq:polemasses}, it seems that a range for $m_c$ of about 1.25--1.55~GeV at 68\% C.L.~and 1.15--1.65~GeV at 90\%~C.L.~would seem reasonable at present.  With improved HERA data on $F_2^c$ to come~\cite{HERAF2charm} this range can hopefully be narrowed in the near future.  Turning now to the bottom mass, the limited information that we obtain from comparison to data, and from the determination of the pole mass given in Eq.~\eqref{eq:polemasses}, suggest that a range of $m_b$ from about 4.65--5.05~GeV at 68\%~C.L.~would seem sensible. As we will see in the next section, LHC cross sections that do not involve explicit $b$-quarks are not very sensitive to the PDF variation with $m_b$, so using our grids at 4.5 and 5~GeV should give a good estimate of the uncertainty due to $m_b$ at 68\%~C.L., and the grids at 4.25 and 5.25~GeV should give a conservative estimate of this source of uncertainty at 90\%~C.L.  As with charm, improved HERA data on $F_2^b$ are likely to limit the allowed spread in future.

If we reach the stage where we become confident that the data are starting to constrain the pole masses to an accuracy clearly better than the renormalon ambiguity, it may become preferable to transform to a scheme where $\overline{\rm MS}$ definitions are used instead, even though the mass is less directly related to a physical variable in this case.  This would be appropriate if we become strongly constrained by data with $Q^2\sim m_h^2$ and data close to the kinematic threshold, $W^2=Q^2(1/x-1)=4m_h^2$ (for neutral-current DIS), where non-perturbative effects and the interplay between the leading-twist power-series and the power corrections become very important, and the ambiguities may be reduced in a different renormalisation scheme.  Indeed, we have already noted in Section 9.2 of Ref.~\cite{Martin:2009iq} that the lowest-$Q^2$ EMC data on $F_2^c$~\cite{Aubert:1982tt} imply a non-perturbative correction to the cross section.

\subsection{Impact on $W$, $Z$ and Higgs production at the Tevatron and LHC}
\begin{table}
  \centering
  (a)\hfill$\,$\\
{\footnotesize
  \begin{tabular}{c|c||r|r|r||r|r|r||r|r|r}
    \hline\hline
    \multicolumn{2}{c||}{Variable $\alpha_S(M_Z^2)$} & \multicolumn{3}{c||}{Tevatron} & \multicolumn{3}{c||}{LHC} & \multicolumn{3}{c}{LHC} \\
    \multicolumn{2}{c||}{} & \multicolumn{3}{c||}{($\sqrt{s}=1.96$ TeV)} & \multicolumn{3}{c||}{($\sqrt{s}=7$ TeV)} & \multicolumn{3}{c}{($\sqrt{s}=14$ TeV)} \\
    \hline
    $m_c$~(GeV) & $m_b$~(GeV) &
    $\delta\sigma^{W}$ & $\delta\sigma^{Z}$ & $\delta\sigma^{H}$ & $\delta\sigma^{W}$ & $\delta\sigma^{Z}$ & $\delta\sigma^{H}$ & $\delta\sigma^{W}$ & $\delta\sigma^{Z}$ & $\delta\sigma^{H}$ \\
    \hline
    1.05 & & -2.6 & -2.8 & +0.4 & -4.1 & -4.6 & -2.4 & -5.1 & -5.5 & -3.8 \\
    1.10 & & -2.2 & -2.4 & +0.2 & -3.5 & -3.9 & -2.1 & -4.3 & -4.7 & -3.3 \\
    1.15 & & -1.8 & -1.9 & +0.1 & -2.9 & -3.3 & -1.8 & -3.6 & -3.9 & -2.8 \\
    1.20 & & -1.4 & -1.5 & +0.1 & -2.3 & -2.6 & -1.5 & -2.8 & -3.1 & -2.3 \\
    1.25 & & -1.0 & -1.1 &  0.0 & -1.7 & -1.9 & -1.2 & -2.1 & -2.3 & -1.7 \\
    1.30 & & -0.7 & -0.7 &  0.0 & -1.1 & -1.3 & -0.8 & -1.4 & -1.5 & -1.2 \\
    1.35 & & -0.3 & -0.4 &  0.0 & -0.6 & -0.6 & -0.4 & -0.7 & -0.8 & -0.6 \\
    {\bf 1.40} & {\bf 4.75} & 0.0 & 0.0 & 0.0 & 0.0 & 0.0 & 0.0 & 0.0 & 0.0 & 0.0 \\
    1.45 & & +0.3 & +0.3 &  0.0 & +0.6 & +0.6 & +0.4 & +0.7 & +0.8 & +0.6 \\
    1.50 & & +0.6 & +0.6 &  0.0 & +1.1 & +1.3 & +0.8 & +1.3 & +1.5 & +1.2 \\
    1.55 & & +0.8 & +0.9 & +0.1 & +1.6 & +1.9 & +1.2 & +2.0 & +2.3 & +1.8 \\
    1.60 & & +1.1 & +1.2 & +0.2 & +2.1 & +2.5 & +1.8 & +2.6 & +3.0 & +2.5 \\
    1.65 & & +1.3 & +1.5 & +0.1 & +2.6 & +3.0 & +2.0 & +3.2 & +3.7 & +2.9 \\
    1.70 & & +1.5 & +1.8 & +0.2 & +3.1 & +3.6 & +2.5 & +3.8 & +4.4 & +3.6 \\
    1.75 & & +1.8 & +2.0 & +0.3 & +3.5 & +4.2 & +2.9 & +4.3 & +5.1 & +4.1 \\
    \hline\hline
  \end{tabular}
}
\\[1cm]
(b)\hfill$\,$\\
{\footnotesize
  \begin{tabular}{c|c||r|r|r||r|r|r||r|r|r}
    \hline\hline
    \multicolumn{2}{c||}{Fixed $\alpha_S(M_Z^2)$} & \multicolumn{3}{c||}{Tevatron} & \multicolumn{3}{c||}{LHC} & \multicolumn{3}{c}{LHC} \\
    \multicolumn{2}{c||}{} & \multicolumn{3}{c||}{($\sqrt{s}=1.96$ TeV)} & \multicolumn{3}{c||}{($\sqrt{s}=7$ TeV)} & \multicolumn{3}{c}{($\sqrt{s}=14$ TeV)} \\
    \hline
    $m_c$~(GeV) & $m_b$~(GeV) &
    $\delta\sigma^{W}$ & $\delta\sigma^{Z}$ & $\delta\sigma^{H}$ & $\delta\sigma^{W}$ & $\delta\sigma^{Z}$ & $\delta\sigma^{H}$ & $\delta\sigma^{W}$ & $\delta\sigma^{Z}$ & $\delta\sigma^{H}$ \\
    \hline
    1.05 & & -1.9 & -2.2 & +3.8 & -2.9 & -3.4 & 0.0 & -3.9 & -4.3 & -1.5 \\
    1.10 & & -1.5 & -1.8 & +3.3 & -2.5 & -2.9 & 0.0 & -3.3 & -3.6 & -1.3 \\
    1.15 & & -1.2 & -1.4 & +2.7 & -2.1 & -2.4 & 0.0 & -2.7 & -3.0 & -1.1 \\
    1.20 & & -0.9 & -1.1 & +2.2 & -1.6 & -1.9 & 0.0 & -2.1 & -2.4 & -0.9 \\
    1.25 & & -0.7 & -0.8 & +1.6 & -1.2 & -1.4 & 0.0 & -1.6 & -1.8 & -0.7 \\
    1.30 & & -0.4 & -0.5 & +1.1 & -0.8 & -1.0 & 0.0 & -1.0 & -1.2 & -0.4 \\
    1.35 & & -0.2 & -0.3 & +0.5 & -0.4 & -0.5 & 0.0 & -0.5 & -0.6 & -0.2 \\
    {\bf 1.40} & {\bf 4.75} & 0.0 & 0.0 & 0.0 & 0.0 & 0.0 & 0.0 & 0.0 & 0.0 & 0.0 \\
    1.45 & & +0.2 & +0.2 & -0.6 & +0.4 & +0.5 & 0.0 & +0.5 & +0.6 & +0.3 \\
    1.50 & & +0.3 & +0.4 & -1.2 & +0.8 & +0.9 & 0.0 & +1.0 & +1.2 & +0.4 \\
    1.55 & & +0.5 & +0.7 & -1.7 & +1.1 & +1.4 & 0.0 & +1.4 & +1.7 & +0.6 \\
    1.60 & & +0.7 & +0.8 & -2.3 & +1.4 & +1.8 & 0.0 & +1.9 & +2.2 & +0.9 \\
    1.65 & & +0.8 & +1.1 & -2.9 & +1.8 & +2.3 & -0.1 & +2.3 & +2.8 & +1.0 \\
    1.70 & & +0.9 & +1.2 & -3.5 & +2.1 & +2.7 & -0.1 & +2.7 & +3.3 & +1.2 \\
    1.75 & & +1.1 & +1.4 & -4.0 & +2.4 & +3.1 & -0.1 & +3.1 & +3.8 & +1.3 \\
    \hline\hline
  \end{tabular}
}
  \caption{\sf (a) Difference (in percent) between the predictions for the $W$, $Z$ and Standard Model Higgs ($M_H = 120$~GeV) NNLO production cross sections at the Tevatron and LHC, calculated using the variable-$m_c$ PDF sets, and the standard MSTW~2008 NNLO predictions with $m_c=1.40$~GeV.  (b) Also shown are the corresponding results for variable $m_c$ with \emph{fixed} $\alpha_S(M_Z^2)$.}
  \label{tab:WZHmc}
\end{table}
\begin{table}
  \centering
{\footnotesize
  \begin{tabular}{c|c||r|r|r||r|r|r||r|r|r}
    \hline\hline
    \multicolumn{2}{c||}{Variable $\alpha_S(M_Z^2)$} & \multicolumn{3}{c||}{Tevatron} & \multicolumn{3}{c||}{LHC} & \multicolumn{3}{c}{LHC} \\
    \multicolumn{2}{c||}{} & \multicolumn{3}{c||}{($\sqrt{s}=1.96$ TeV)} & \multicolumn{3}{c||}{($\sqrt{s}=7$ TeV)} & \multicolumn{3}{c}{($\sqrt{s}=14$ TeV)} \\
    \hline
    $m_c$~(GeV) & $m_b$~(GeV) &
    $\delta\sigma^{W}$ & $\delta\sigma^{Z}$ & $\delta\sigma^{H}$ & $\delta\sigma^{W}$ & $\delta\sigma^{Z}$ & $\delta\sigma^{H}$ & $\delta\sigma^{W}$ & $\delta\sigma^{Z}$ & $\delta\sigma^{H}$ \\
    \hline
    & 4.00 & -0.2 & -0.1 & -0.2 & -0.4 & -0.1 & -0.4 & -0.5 &  0.0 & -0.5 \\
    & 4.25 & -0.1 & -0.1 & -0.1 & -0.3 &  0.0 & -0.3 & -0.3 &  0.0 & -0.3 \\
    & 4.50 & -0.1 &  0.0 & -0.1 & -0.1 &  0.0 & -0.1 & -0.1 &  0.0 & -0.1 \\
    {\bf 1.40} & {\bf 4.75} & 0.0 & 0.0 & 0.0 & 0.0 & 0.0 & 0.0 & 0.0 & 0.0 & 0.0 \\
    & 5.00 &  0.0 &  0.0 &  0.0 & +0.1 &  0.0 & +0.1 & +0.1 &  0.0 & +0.1 \\
    & 5.25 & +0.1 &  0.0 & +0.1 & +0.2 &  0.0 & +0.2 & +0.2 &  0.0 & +0.2 \\
    & 5.50 & +0.1 &  0.0 & +0.1 & +0.3 &  0.0 & +0.2 & +0.3 & -0.1 & +0.3 \\
    \hline\hline
  \end{tabular}
}
  \caption{\sf Difference (in percent) between the predictions for the $W$, $Z$ and Standard Model Higgs ($M_H = 120$~GeV) NNLO production cross sections at the Tevatron and LHC, calculated using the variable-$m_b$ PDF sets, and the standard MSTW~2008 NNLO predictions with $m_b=4.75$~GeV.  Here $\alpha_S(M_Z^2)$ is a free parameter in each fit, although the correlation with $m_b$ is negligible, see Table~\ref{tab:mb}.}
  \label{tab:WZHmb}
\end{table}

\begin{figure}
  \centering
  (a)\hfill$\,$\\
  \includegraphics[width=0.8\textwidth]{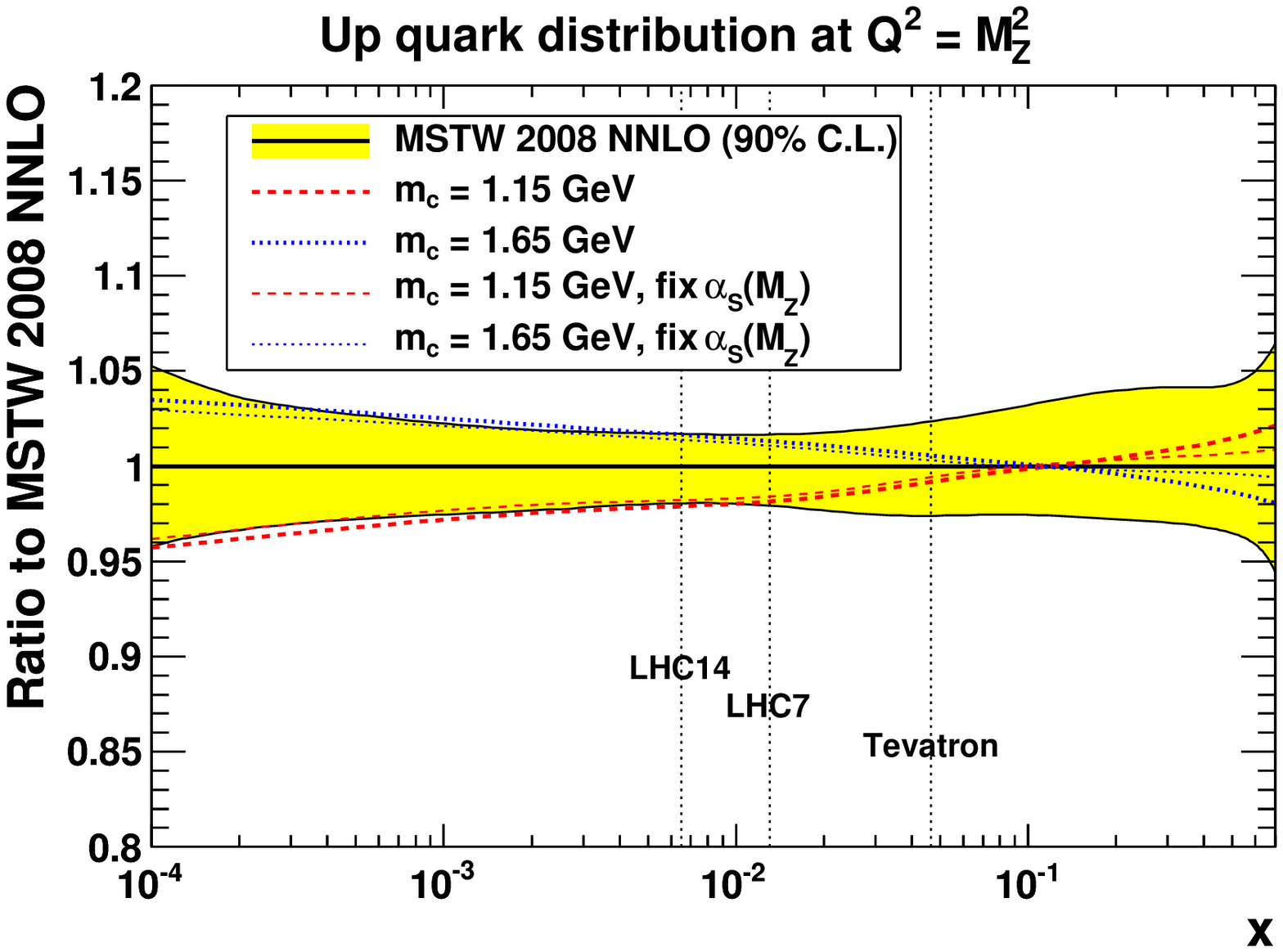}\\
  (b)\hfill$\,$\\
  \includegraphics[width=0.8\textwidth]{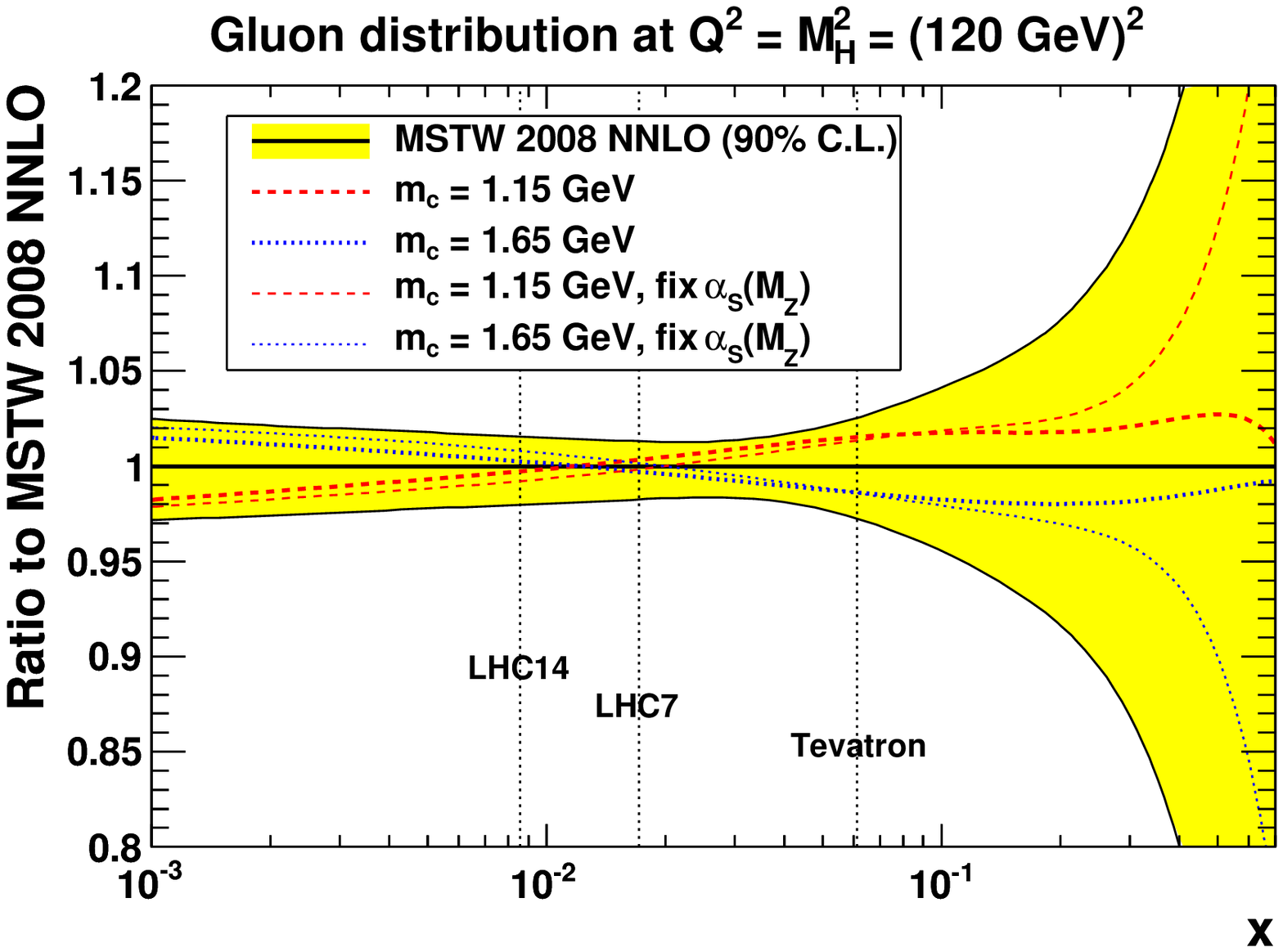}
  \caption{\sf $m_c$ dependence of the NNLO (a) up-quark and (b) gluon distributions at the scales relevant for $Z$ and Higgs boson ($M_H=120$ GeV) production, respectively.  Also shown is the $m_c$ dependence when $\alpha_S(M_Z^2)$ is held fixed.  The values of $x=M_{Z,H}/\sqrt{s}$ probed at central rapidity (and $p_T^{Z,H}=0$) are indicated for the Tevatron ($\sqrt{s} = 1.96$ TeV) and LHC ($7$~TeV and $14$~TeV).}
  \label{fig:mcdepnnlo}
\end{figure}

Tables~\ref{tab:WZHmc} and \ref{tab:WZHmb} show how the predictions for the ``standard candle'' NNLO $W$ and $Z$ cross sections at the Tevatron and LHC change using the PDFs obtained when the charm- and bottom-quark masses are varied in the global fit.  The cross sections are calculated in the 5-flavour ZM-VFNS as described in Section~15 of Ref.~\cite{Martin:2009iq}, e.g.~using the PDG 2008~\cite{Amsler:2008zzb} electroweak parameters.  These changes are not due primarily to the changes in the heavy-quark PDFs themselves, which would in any case be in the opposite direction.  Rather they are due to the changes in the light quarks, which evolve slightly more rapidly at small $x$ to compensate for the slower turn-on of the heavy-flavour contribution to the structure functions when the heavy-quark masses are increased.  In the free coupling case this is achieved mainly by an increase in $\alpha_S$, while for fixed coupling it occurs from an increase in the small-$x$ input gluon distribution.  Additionally, the input sea quarks, mainly the less well-constrained strange quark distributions, also increase with increasing $m_h$ to compensate for the suppressed heavy-flavour contribution to small-$x$ structure functions.  At the Tevatron, varying the charm mass by $\pm 0.15$ GeV from its default MSTW~2008 value leads to fairly small (${\cal O}(1\%)$ or less) changes in the $W$ and $Z$ cross sections, with heavier values of the charm mass giving slightly larger cross sections.  The effects are a little more pronounced at the LHC: the cross sections vary by about $\pm 2\%$ for $\sqrt{s}=14$~TeV.  These changes reflect the behaviour of the up-quark distribution shown in Fig.~\ref{fig:mcdepnnlo}(a) at the relevant $x$ values (indicated).  The sensitivity to the value of $m_c$ of the $W$ and $Z$ cross sections at the LHC has previously been noticed in Ref.~\cite{CooperSarkar:2010ik}.

In Tables~\ref{tab:WZHmc} and \ref{tab:WZHmb} we also give the changes to the cross section for the production via gluon--gluon fusion through a top-quark loop of a Standard Model Higgs boson of mass 120 GeV, calculated in the 5-flavour ZM-VFNS as described in Section 6.2 of Ref.~\cite{Martin:2009bu}.  From Table~\ref{tab:WZHmc} we see that the variation of the Higgs cross sections with $m_c$ is different depending on whether $\alpha_S$ is (a)~allowed to vary or (b)~kept fixed.  This can be understood from the behaviour of the gluon distribution as $m_c$ is varied, see Fig.~\ref{fig:mcdepnnlo}(b).  When $\alpha_S$ is fixed, the change in the Higgs cross section reflects the change in the gluon distribution at the relevant $x$ values (indicated).  Thus the \{14 TeV LHC, 7 TeV LHC, Tevatron\} Higgs cross sections are \{correlated, uncorrelated, anticorrelated\} with $m_c$.  On the other hand, for variable $\alpha_S$, the additional correlation of $\alpha_S$ with $m_c$ (see Table~\ref{tab:mc}) results in \{14 TeV LHC, 7 TeV LHC, Tevatron\} Higgs cross sections that are \{correlated, correlated, almost uncorrelated\} with $m_c$, i.e.~at the Tevatron the anticorrelation of the gluon distribution largely cancels the correlation of $\alpha_S$.

From Table~\ref{tab:WZHmb} we see that the $W$, $Z$ and Higgs production cross sections are much less dependent on the value of $m_b$.   As seen from Table~\ref{tab:mb}, the correlation between $m_b$ and $\alpha_S(M_Z^2)$ is negligible, so we do not explicitly show results with fixed $\alpha_S(M_Z^2)$ since they would be almost the same as in Table~\ref{tab:WZHmb}.  In all cases, for both varying-$m_c$ and varying-$m_b$, the cross-section ratios $\sigma^W/\sigma^Z$ (and also $\sigma^{W^+}/\sigma^{W^-}$ at the LHC) are almost completely unaffected.

\begin{table}
  \centering
  \begin{tabular}{l|c|c|c}
    \hline\hline
    Tevatron, $\sqrt{s} = 1.96$ TeV & $B_{\ell\nu} \cdot \sigma^W$ & $B_{\ell^+\ell^-}\cdot\sigma^Z$ & $\sigma^H$ \\
    \hline
    Central value & 2.747~nb & 0.2507~nb & 0.9550~pb \\ \hline
    PDF only uncertainty & $^{+1.8\%}_{-1.5\%}$ & $^{+1.9\%}_{-1.6\%}$ & $^{+3.1\%}_{-3.3\%}$ \\
    PDF+$\alpha_S$ uncertainty & $^{+2.2\%}_{-1.7\%}$ & $^{+2.2\%}_{-1.8\%}$ & $^{+5.4\%}_{-4.8\%}$ \\ \hline
    PDF+$\alpha_S$+$m_{c,b}$ uncertainty & $^{+2.3\%}_{-1.8\%}$ & $^{+2.3\%}_{-2.0\%}$ & $^{+5.6\%}_{-5.1\%}$ \\
    \hline\hline\multicolumn{4}{c}{}\\\hline\hline
    LHC, $\sqrt{s} = 7$ TeV & $B_{\ell\nu} \cdot \sigma^W$ & $B_{\ell^+\ell^-}\cdot\sigma^Z$ & $\sigma^H$ \\
    \hline
    Central value & 10.47~nb & 0.958~nb & 15.50~pb \\ \hline
    PDF only uncertainty & $^{+1.7\%}_{-1.6\%}$ & $^{+1.7\%}_{-1.5\%}$ & $^{+1.1\%}_{-1.6\%}$ \\
    PDF+$\alpha_S$ uncertainty & $^{+2.5\%}_{-1.9\%}$ & $^{+2.5\%}_{-1.9\%}$ & $^{+3.7\%}_{-2.9\%}$ \\ \hline
    PDF+$\alpha_S$+$m_{c,b}$ uncertainty & $^{+2.7\%}_{-2.2\%}$ & $^{+2.9\%}_{-2.4\%}$ & $^{+3.7\%}_{-2.9\%}$ \\
    \hline\hline\multicolumn{4}{c}{}\\\hline\hline
    LHC, $\sqrt{s} = 14$ TeV & $B_{\ell\nu} \cdot \sigma^W$ & $B_{\ell^+\ell^-}\cdot\sigma^Z$ & $\sigma^H$ \\
    \hline
    Central value & 21.72~nb & 2.051~nb & 50.51~pb \\ \hline
    PDF only uncertainty & $^{+1.7\%}_{-1.7\%}$ & $^{+1.7\%}_{-1.6\%}$ & $^{+1.0\%}_{-1.6\%}$ \\
    PDF+$\alpha_S$ uncertainty & $^{+2.6\%}_{-2.2\%}$ & $^{+2.6\%}_{-2.1\%}$ & $^{+3.6\%}_{-2.7\%}$ \\ \hline
    PDF+$\alpha_S$+$m_{c,b}$ uncertainty & $^{+3.0\%}_{-2.7\%}$ & $^{+3.1\%}_{-2.8\%}$ & $^{+3.7\%}_{-2.8\%}$ \\
    \hline\hline
  \end{tabular}
  \caption{\sf NNLO predictions for $W$, $Z$ and Higgs ($M_H=120$~GeV) total cross sections at the Tevatron, $7$~TeV LHC and $14$~TeV LHC, with PDF uncertainties only~\cite{Martin:2009iq}, with the combined ``PDF+$\alpha_S$'' uncertainty~\cite{Martin:2009bu}, then finally also including the uncertainty due to $m_c$ and $m_b$.  The $68\%$ C.L.~uncertainties are given in all cases.  We take $\mu_R=\mu_F=M_{W,Z,H}$.}
  \label{tab:wzhtot}
\end{table}
In order to assess the {\em total} uncertainty on a hadron collider cross section arising from variations in PDFs, $\alpha_S$ and the heavy-quark masses, we need to define a prescription for combining the latter with the former. Our recommendation is to vary $m_c$ in the range $m_c = 1.40\pm 0.15$~GeV at 68\%~C.L.~and $\pm 0.25$~GeV at 90\%~C.L., centred on the default value for fixed $\alpha_S$, and to vary $m_b$ in the range $m_b = 4.75\pm0.25$~GeV at 68\%~C.L.~and $\pm0.50$~GeV at 90\%~C.L., and then to separately add the resulting cross-section variations in quadrature with the usual ``PDF+$\alpha_S$'' uncertainty.  Of course, this prescription does not account for all possible correlations between PDFs, $\alpha_S$ and $m_{c,b}$, but it should be a sufficiently good approximation.  The range of $m_c$ values is based on the slightly contrasting pulls of the pole mass determination from the $\overline{\rm MS}$ conversion and our fit.  As an example, in Table~\ref{tab:wzhtot} we show the increase in uncertainty on the $W$, $Z$ and Higgs total cross sections due to $m_c$ and $m_b$ variations, compared to the ``PDF only''~\cite{Martin:2009iq} and ``PDF+$\alpha_S$''~\cite{Martin:2009bu} uncertainties.

\begin{table}
  \centering
  \begin{tabular}{l|l|l}
    \hline\hline
    PDF set & $m_c$~(GeV) & $m_b$~(GeV) \\
    \hline
    MSTW08~\cite[this work]{Martin:2009iq} & $1.40\pm0.15$ & $4.75\pm0.25$ \\
    MRST06~\cite{Martin:2007bv} & $1.43$ & $4.30$ \\
    CT10~\cite{Lai:2010vv} & $1.30$ & $4.75$ \\
    CTEQ6.6~\cite{Nadolsky:2008zw} & $1.30$ & $4.50$ \\
    NNPDF2.0~\cite{Ball:2010de} & $\sqrt{2}$ & $4.30$ \\
    GJR08/JR09~\cite{Gluck:2007ck,JimenezDelgado:2008hf} & $1.30$ & $4.20$ \\
    HERAPDF1.0~\cite{Aaron:2009wt} & $1.40^{+0.20}_{-0.05}$ & $4.75^{+0.25}_{-0.45}$ \\
    ABKM09~\cite{Alekhin:2009ni} & $1.50\pm0.10$ & $4.50\pm0.50$ \\
    \hline\hline
  \end{tabular}
  \caption{\sf Values of $m_c$ and $m_b$ used in various PDF fits.  The $1\sigma$ uncertainties are given only for the three PDF groups who attempt to account for uncertainties on $m_{c,b}$ in their analyses.}
  \label{tab:mhcompare}
\end{table}
In Table~\ref{tab:mhcompare} we compare the values of the heavy-quark masses, $m_c$ and $m_b$, used in the MRST/MSTW analyses, with the values taken in the most recent public PDF fits of other groups.  The other two \emph{global} PDF fitting groups, CTEQ~\cite{Lai:2010vv,Nadolsky:2008zw} and NNPDF~\cite{Ball:2010de}, do not attempt to quantify the uncertainty coming from $m_{c,b}$, neither do the fits of ``dynamical'' PDFs by GJR/JR~\cite{Gluck:2007ck,JimenezDelgado:2008hf}.  The HERAPDF1.0 fit~\cite{Aaron:2009wt} uses the same central values as MSTW08~\cite{Martin:2009iq}, but additional PDF fits are also provided with $m_c=\{1.35,1.65\}$~GeV and $m_b=\{4.3,5.0\}$~GeV, and this additional model uncertainty is recommended to be added in quadrature with the other uncertainties.  The ABKM09 analysis~\cite{Alekhin:2009ni} uses fixed values of $m_c=1.5$~GeV and $m_b=4.5$~GeV to determine the central fit.  However, additional pseudo-measurements of $m_{c,b}$ are then added, with values given in the last line of Table~\ref{tab:mhcompare}, and $m_c$ and $m_b$ are taken as free parameters to calculate the covariance matrix used for the final error propagation.  This means that each of the public eigenvector PDF sets will be associated with different values of $m_c$ and $m_b$, but these values are not readily accessible.

\section{3-flavour and 4-flavour scheme parton distributions} \label{sec:FFNS}
As well as looking at the variation in the parton distributions as a function of the heavy-quark masses, it is also interesting to consider the PDF sets obtained in the framework of a different maximum number of active quark flavours. Hence, in this section we will consider our PDFs when charm becomes an active parton but bottom does not---the 4-flavour scheme---and when both the charm and bottom quarks only appear in the final state---the 3-flavour scheme.  We have argued on various occasions (e.g.~in Section 4 of Ref.~\cite{Martin:2009iq}) that the use of a GM-VFNS, i.e.~with up to five active quarks (or even six if we include top), is preferable.  However, there are cases where the cross section has only been calculated with finite mass dependence for the fixed flavour number case; see, for example, the \textsc{hqvdis} program~\cite{Harris:1995tu,Harris:1997zq} for details of final states in heavy-quark production in DIS.  For this reason we make available sets of 3FS and 4FS distributions with a variety of both charm and bottom masses.

\subsection{Obtaining 3-flavour and 4-flavour scheme parton distributions}
It might be thought preferable to obtain these lower active quark number PDF sets from a fit performed using FFNS scheme coefficient functions. However, as argued in Refs.~\cite{Martin:2006qz,Thorne:2008xf}, it is not actually so obvious that this is the case. This is largely because rather few of the data sets included in a truly global PDF fit can be kept, even at NLO, in a fit using the FFNS, due to lack of the full coefficient functions (even charged-current DIS coefficients are not known with full mass dependence at order $\alpha_S^2$ except for $Q^2 \gg m_h^2$~\cite{Buza:1997mg}). Hence, the central values of the PDFs are likely to be influenced as much by the lack of data as by the change of scheme. However, it is also a consideration that the lack of resummation of the large logarithms in $Q^2/m_h^2$ potentially affects the stability of the fit compared to the presumably more stable GM-VFNS fit.  Ultimately a correct GM-VFNS will provide results very similar to the FFNS near to the transition points $Q^2=m_h^2$ anyway, so we deem it best to simply obtain the 3FS and 4FS PDFs from the inputs for the full fits performed using the GM-VFNS. 

\begin{figure}
  \centering
  \includegraphics[width=0.8\textwidth]{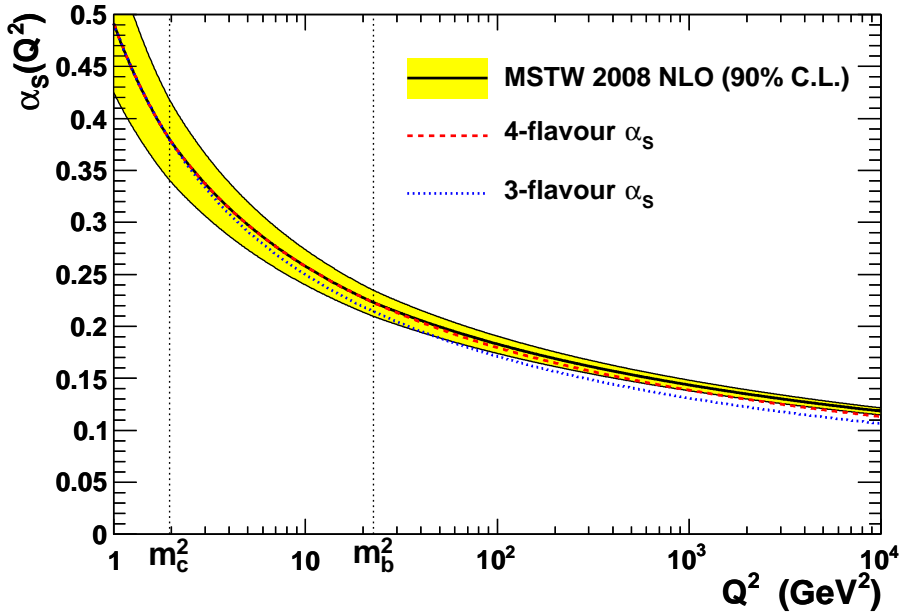}
  \caption{\sf The NLO strong coupling $\alpha_S(Q^2)$ versus $Q^2$, in the default 5FS with the $90\%$~C.L.~experimental uncertainty, and in the 4FS and 3FS, taking the same input value of $\alpha_S(Q_0^2=1\,{\rm GeV}^2)$.}
  \label{fig:alphas}
\end{figure}

\begin{figure}
  \centering
  (a)\hfill$\,$\\
  \includegraphics[width=0.8\textwidth]{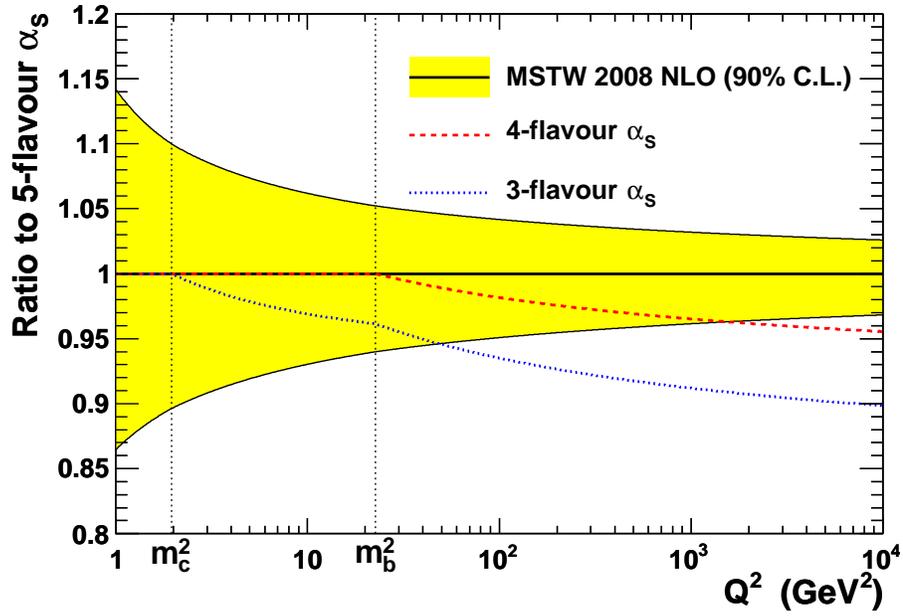}\\
  (b)\hfill$\,$\\
  \includegraphics[width=0.8\textwidth]{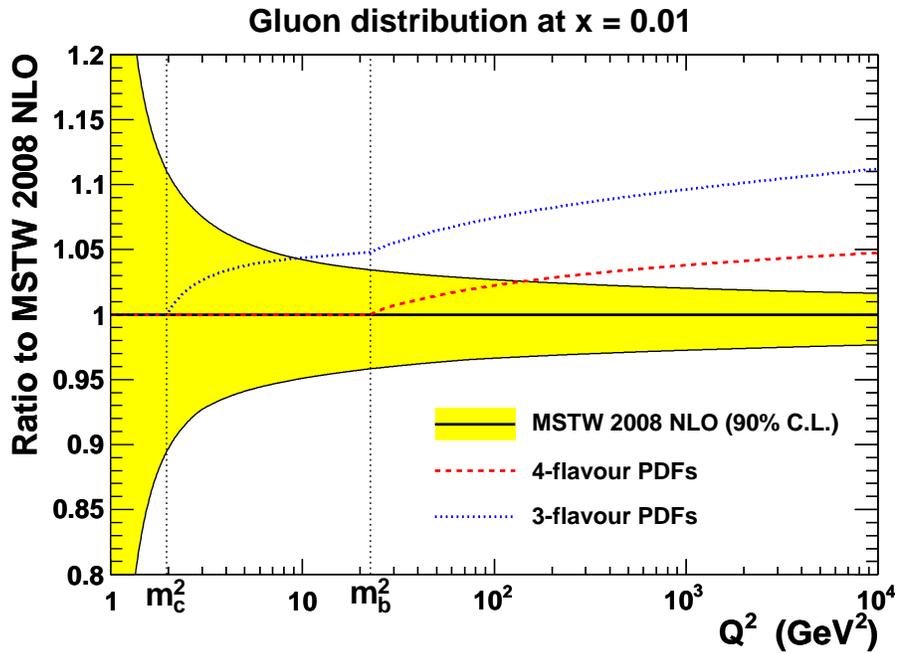}
  \caption{\sf Ratio to the 5FS values of (a)~strong coupling $\alpha_S$ and (b)~gluon distribution at $x=0.01$, versus $Q^2$ at NLO in the 4FS and 3FS, compared to the $90\%$~C.L.~experimental uncertainty bands.}
  \label{fig:ratioasg}
\end{figure}

When obtaining the 3FS and 4FS PDFs it is vital to make a self-consistent definition of the strong coupling constant $\alpha_S$.  It is generally the case that coefficient functions calculated in the 3- or 4-flavour schemes are made in a renormalisation scheme where the contribution of the heavy quark decouples and the coupling itself does not include the heavy quark as an active flavour.  On this basis we define the PDFs using this definition of the coupling in the splitting functions. It is certainly possible to use a different definition of the coupling, but this must be applied universally, in both the PDFs and the coefficient functions. As illustrated in Ref.~\cite{Martin:2006qz}, the error, made from not doing so, can be a few percent.  Indeed, the change in the coupling obtained by altering the number of heavy flavours is quite dramatic. In our default NLO fit we start with a value $\alpha_S(Q_0^2) = 0.49128$ (with $Q_0^2=1$~GeV$^2$), which for a variable flavour number results in $\alpha_S^{(5)}(M_Z^2) = 0.12018$. Restricting ourselves to a maximum of four active flavours, the same boundary condition at $Q_0^2$ gives $\alpha_S^{(4)}(M_Z^2) = 0.11490$, and for three active flavours gives $\alpha_S^{(3)}(M_Z^2) = 0.10809$.  This is illustrated in Fig.~\ref{fig:alphas}, and more clearly in Fig.~\ref{fig:ratioasg}(a), which shows the ratio of the 3- and 4-flavour $\alpha_S$ to the 5-flavour one.

\begin{figure}
  \centering
  \includegraphics[width=0.8\textwidth]{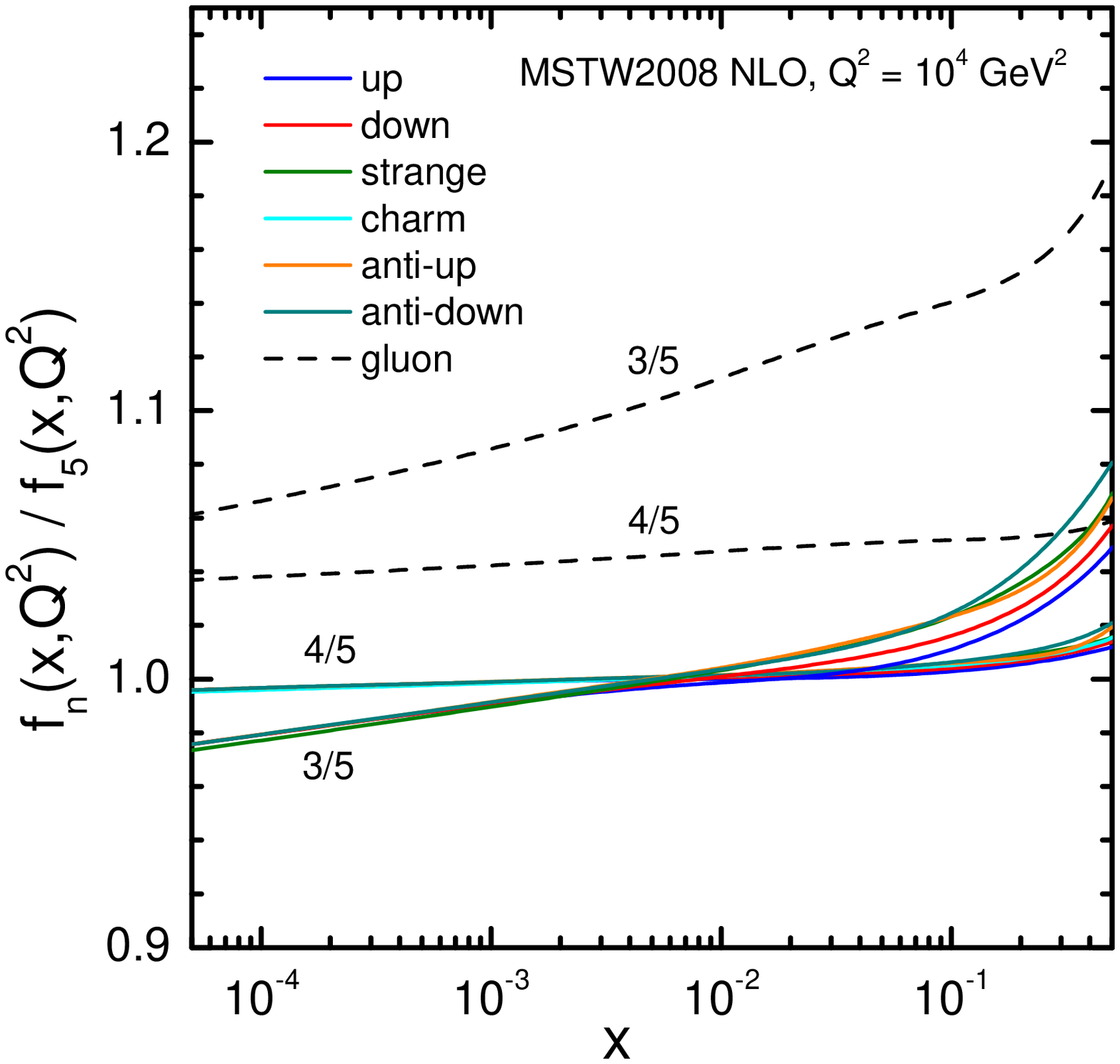}
  \caption{\sf Ratio of 3FS and 4FS NLO PDFs to the standard 5FS NLO PDFs, at $Q^2 = 10^4$~GeV$^2$.}
  \label{fig:partons345}
\end{figure}
We make available both the 3FS and 4FS NLO PDF sets for the full variety of fits with varying charm and bottom masses discussed in the previous section, including the case of varying $m_c$ with fixed $\alpha_S(M_Z^2)$.  We do not vary the quark mass at LO since the quality of the fit is already poor at LO.  However, we do provide the 3FS and 4FS PDFs for the default masses at LO since these may be useful in some Monte Carlo event generators.  We also make available the full range of 3FS and 4FS sets at NNLO, although the application of these is distinctly limited at present due to the lack of cross sections with full mass dependence calculated at this order. We will comment more on this in the following section. The ratios of the 3FS and 4FS NLO PDFs to those in the 5FS are shown in Fig.~\ref{fig:partons345} at $Q^2=10^4$~GeV$^2$.  In both cases, the gluon is larger in the 4FS and, particularly, in the 3FS, due to the fact that it splits into a smaller number of quark--antiquark pairs; see also Fig.~\ref{fig:ratioasg}(b).  The effect of the increased growth of the gluon distribution is countered exactly for the leading term in $\alpha_S(Q^2)\ln(Q^2/m_h^2)$ by the quicker decrease of the coupling.  This is illustrated in Fig.~\ref{fig:ratioasg}, where the change in the gluon distribution comparing different flavour numbers is, to a good approximation, the inverse of that for $\alpha_S$.  This behaviour is only violated by higher-order corrections, so the gluon-driven change in the light quarks at small $x$ is rather minimal.  At higher $x$, the onset of the smaller coupling in the lower flavour number schemes means a slower evolution and consequently larger light-quark distributions, as seen in Fig.~\ref{fig:partons345}.

Finally, we also make available new eigenvector PDF sets for the 3- and 4-flavour PDFs using the default quark masses at LO, NLO and NNLO.  These are evolved from the saved PDF parameters at $Q_0^2=1$~GeV$^2$ for the default MSTW fits, i.e.~the ``dynamical'' tolerance values, $T=(\Delta\chi^2_{\rm global})^{1/2}$, are those determined from the 5-flavour fit~\cite{Martin:2009iq}.  For the first time, this will allow PDF uncertainties to be consistently included in a 3FS or 4FS calculation.  We have not yet generated the additional eigenvector PDF sets with varying $\alpha_S$ values needed for the ``PDF+$\alpha_S$'' uncertainty calculation~\cite{Martin:2009bu}, although these could be provided at a future point if necessary.  Note, however, that 3FS or 4FS calculations generally have large higher-order corrections, and an associated large (renormalisation and factorisation) scale dependence, which is common in processes with multiple hard scales.  The theory uncertainty due to neglected higher-order corrections is therefore likely to dominate over any PDF or ``PDF+$\alpha_S$'' uncertainty.  Precise calculations of the latter quantities are therefore relatively less important than in typical 5FS calculations where higher-order corrections are more readily available.

\subsection{Comparison of $Z$ total cross-section predictions in 4FS and 5FS} \label{sec:Ztot4FS}
We have already noted in Section~2 that while the parton distributions of a $n$FS, with appropriate coefficient functions, can give an adequate description of the production of an $(n+1)$th parton near threshold, the accuracy of the perturbative expansion becomes increasingly unreliable as the scale of the hard scattering process, $Q^2$, increases above $m_h^2$. In this section we illustrate this by considering the total $Z$ cross section at hadron colliders. In particular, we compare the cross-section predictions in the 5FS ($\equiv$ MSTW~2008~\cite{Martin:2009iq}) and 4FS defined previously. Since in this case $Q^2 = M_Z^2 \gg m_b^2$, we would expect to see quantitative differences between the predictions due to the resummed $[\alpha_S \log(M_Z^2/m_b^2)]^n$ terms which are implicit in the 5FS, via the evolution of the $b$-quark PDF and $\alpha_S^{(5)}$, but absent from the 4FS.  For reasons which will become apparent below, we consider the $Z$ cross section at both NLO and NNLO.

In the 4FS the $b$ quarks are not considered as partons in the initial state, but contribute to the $Z$ cross section via real and virtual contributions which first appear at ${\cal O}(\alpha_S^2)$ in perturbation theory.  Therefore the only difference in the predicted cross sections at NLO are (i) explicit $b$-parton contributions $b + \bar b \to Z ( + g)$ and $g + b (\bar b) \to Z + b (\bar b)$ which only contribute in the $n_f = 5$ scheme (5FS), (ii) small differences in the light-quark and gluon distributions arising from the slightly different evolution in the two schemes, see Fig.~\ref{fig:partons345}, and (iii) the difference in the values of $\alpha_S^{(4)}(M_Z^2)$ and $\alpha_S^{(5)}(M_Z^2)$, see Figs.~\ref{fig:alphas} and \ref{fig:ratioasg}(a), which affect the size of the NLO $K$-factor.  We would expect the Tevatron $Z$ cross sections to be more similar than those at the LHC, since the $b$-quark contributions to the 5FS cross section are much smaller at the lower collider energy.  We do not explicitly include PDF and/or $\alpha_S$ uncertainties on the calculated cross sections, as these are presumed to be more or less the same for the 4FS and 5FS calculations.  Contributions from top quarks are also not included, as these have been shown to be very small in Ref.~\cite{Rijken:1995gi}, and in any case should be added to both the 4FS and 5FS cross sections.

\begin{table}
  \centering
  \begin{tabular}{l|c|c|c}
    \hline\hline
    Tevatron, $\sqrt{s} = 1.96$ TeV &  $B\cdot\sigma_{\rm NLO}^Z(4{\rm FS})$ (nb) & $B\cdot\sigma_{\rm NLO}^Z(5{\rm FS})$ (nb)
   & $B\cdot\sigma_{\rm NLO}^Z(5{\rm FS},b)$ (nb) \\ \hline
$\sigma_0^Z$   & 0.1989 & 0.1990 & 0.0012\\
$\sigma_1^Z$   & 0.0413 & 0.0436 & -0.0002\\
 \hline
total & 0.2402  & 0.2426 & 0.0010\\ 
    \hline\hline\multicolumn{4}{c}{}\\\hline\hline
    LHC, $\sqrt{s} = 7$ TeV &  $B\cdot\sigma_{\rm NLO}^Z(4{\rm FS})$ (nb) & $B\cdot\sigma_{\rm NLO}^Z(5{\rm FS})$ (nb)
   & $B\cdot\sigma_{\rm NLO}^Z(5{\rm FS},b)$ (nb) \\ \hline
$\sigma_0^Z$   & 0.7846 & 0.8023 & 0.0205 \\
$\sigma_1^Z$   & 0.1206 & 0.1285 & -0.0020\\
 \hline
total &  0.9052 & 0.9308 & 0.0185\\ 
    \hline\hline\multicolumn{4}{c}{}\\\hline\hline
    LHC, $\sqrt{s} = 14$ TeV &  $B\cdot\sigma_{\rm NLO}^Z(4{\rm FS})$ (nb) & $B\cdot\sigma_{\rm NLO}^Z(5{\rm FS})$ (nb)
 & $B\cdot\sigma_{\rm NLO}^Z(5{\rm FS},b)$ (nb) \\  \hline
$\sigma_0^Z$   & 1.6922 & 1.7545 & 0.0656\\
$\sigma_1^Z$   & 0.2303 & 0.2465 & -0.0050\\
 \hline
total & 1.9225  & 2.0009 & 0.0601\\ 
\hline\hline
  \end{tabular}
  \caption{\sf NLO predictions for the total $Z$ cross section (multiplied by leptonic branching ratio $B$) at the Tevatron and LHC using MSTW~2008 NLO PDFs~\cite{Martin:2009iq} as input, broken down into the $\alpha_S^n$ ($n=0,1$) contributions, with $\{q=u,d,s,c;\; \alpha_S^{(4)};\; $4-flavour MSTW~2008 NLO PDFs$\}$ in the 4FS calculation and $\{q=u,d,s,c,b;\; \alpha_S^{(5)};\; $5-flavour MSTW~2008 NLO PDFs$\}$ in the 5FS calculation.  The final column gives the contribution to the 5FS cross sections from processes where the $Z$ couples directly to $b$ quarks.}
  \label{tab:z01nlotot}
\end{table}
The results at NLO are shown in Table~\ref{tab:z01nlotot}.\footnote{The 5FS cross sections are identical to those reported in Ref.~\cite{Martin:2009iq}.}  The 4FS cross section is smaller by $1.0\%$, $2.8\%$ and $3.9\%$ at the Tevatron, LHC(7 TeV) and LHC(14 TeV) respectively.  The ${\cal O}(\alpha_S^1)$ contributions differ by more than the ${\cal O}(\alpha_S^0)$ contributions, due to the $\sim5\%$ differences in the $\alpha_S(M_Z^2)$ values in the two schemes.  Also shown in Table~\ref{tab:z01nlotot} (final column) are the contributions to the 5FS cross sections from the $b + \bar b \to Z ( + g)$ and $g + b (\bar b) \to Z + b (\bar b)$  processes, which evidently account for $0.4\%$, $2.0\%$ and $3.0\%$ of the total at the Tevatron, LHC(7 TeV) and LHC(14 TeV) respectively.  Thus these contributions account for the bulk of the differences in the 4FS and 5FS cross sections at the LHC energies.

As noted above, when making a comparison between the 4FS and 5FS calculations at NNLO, we must include in the former the explicit real and virtual $b$-quark contributions which first appear at ${\cal O}(\alpha_S^2)$ in perturbation theory.  When calculated with a non-zero $b$-quark mass\footnote{For consistency, we use $m_b = 4.75$~GeV, the same value used in the 5FS~\cite{Martin:2009iq} to generate the $b$-quark PDFs via the DGLAP equation.} these contributions are finite, and simply add to the 4-flavour contributions.
\begin{figure}
  \begin{center}
    \includegraphics[width=0.7\textwidth]{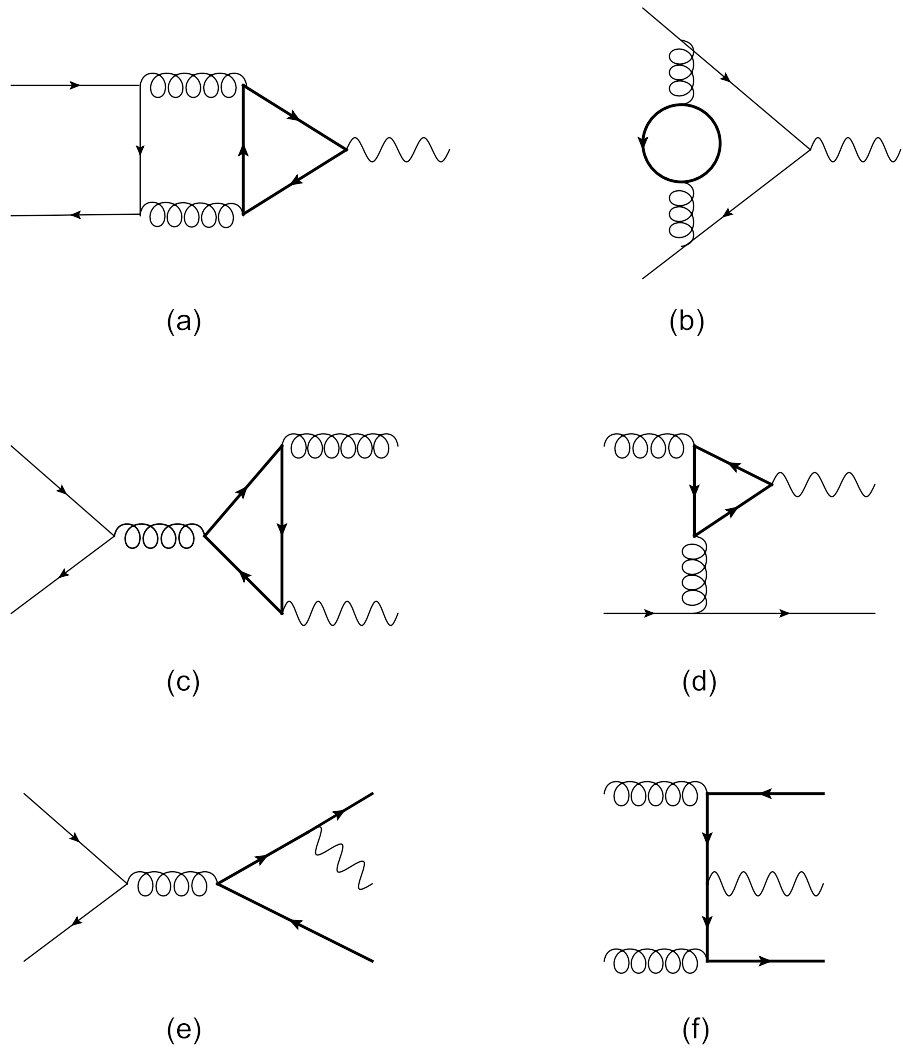}
    \caption{\sf Sample Feynman diagrams for the various (a,b,c,d) virtual and (e,f) real $b$-quark contributions to the $Z$ cross section.  The $b$-quarks are shown as the thicker fermion lines.}
    \label{fig:diagrams}
  \end{center}
\end{figure}
Sample Feynman diagrams for the various real and virtual $b$-quark contributions are shown in Fig.~\ref{fig:diagrams}, and analytic expressions are given in Refs.~\cite{Rijken:1995gi,Gonsalves:1991qn}.  We summarise these in Appendix~\ref{sec:appendix}, and derive small-mass (i.e.~$m_b^2/M_Z^2 \ll 1$) expansions which are useful in practice.
\begin{table}
  \centering
  \begin{tabular}{l|r|r|r}
    \hline\hline
subprocess & $\Delta_b\sigma^Z$ (Tevatron) & $\Delta_b\sigma^Z$ (LHC, 7 TeV) & $\Delta_b\sigma^Z$ (LHC, 14 TeV)  \\
\hline
$q + \bar q \to Z$ &  $5.230 \times 10^{-6}$ & $-2.124 \times 10^{-5}$ & $-6.440 \times 10^{-5}$ \\ \hline
$q + \bar q \to Z + g$ &  $4.901 \times 10^{-5}$ & $6.185 \times 10^{-5}$ & $9.701 \times 10^{-5}$ \\ \hline
$q(\bar q) + g  \to Z + q(\bar q)$ &  $-2.862 \times 10^{-5}$ & $-1.456 \times 10^{-4}$ & $-2.632 \times 10^{-4}$ \\ \hline
$q + \bar q \to Z + b + \bar b$ &  $3.754 \times 10^{-4}$  & $1.450 \times 10^{-3}$ & $3.382 \times 10^{-3}$ \\ \hline
$ g + g \to Z + b + \bar b$ &  $2.090 \times 10^{-4}$  & $5.287 \times 10^{-3}$ & $1.997 \times 10^{-2}$ \\ \hline
total &  $6.100 \times 10^{-4}$  &    $6.632 \times 10^{-3}$ & $2.312 \times 10^{-2}$\\
    \hline\hline
  \end{tabular}
  \caption{\sf Additional ${\cal O}(\alpha_S^2)$ contributions to the total $Z$ 4FS NNLO cross section in nb (multiplied by leptonic branching ratio) at the Tevatron and LHC arising from real and virtual $b$-quark processes.}
  \label{tab:ztotb}
\end{table}
The results are presented in Table~\ref{tab:ztotb}, for Tevatron and ($\sqrt{s} = 7, 14$~TeV) LHC energies.  Evidently the $2\to 3$ process contributions, Eq.~(\ref{eq:3}), are by far the most dominant, and especially so at the LHC where they are two orders of magnitude larger then the rest of the ${\cal O}(\alpha_S^2)$  contributions combined. 
\begin{table}
  \centering
  \begin{tabular}{l|c|c|c}
    \hline\hline
    Tevatron, $\sqrt{s} = 1.96$ TeV &  $B\cdot\sigma_{\rm NNLO}^Z(4{\rm FS})$ (nb) & $B\cdot\sigma_{\rm NNLO}^Z(5{\rm FS})$ (nb)
  & $B\cdot\sigma_{\rm NNLO}^Z(5{\rm FS},b)$ (nb)\\  \hline
$\sigma_0^Z$   & 0.2013 & 0.2016 & 0.0012 \\
$\sigma_1^Z$   & 0.0409 & 0.0431 & -0.0002\\
$\sigma_2^Z$   & 0.0063 & 0.0060 & -0.0003\\ \hline
total & 0.2485  & 0.2507 & 0.0008 \\ \hline
$\Delta_b\sigma^Z$ & 0.0006  & $-$ & \\ \hline
total + $\Delta_b\sigma^Z$  & 0.2491  & 0.2507 & \\
    \hline\hline\multicolumn{4}{c}{}\\\hline\hline
    LHC, $\sqrt{s} = 7$ TeV &  $B\cdot\sigma_{\rm NNLO}^Z(4{\rm FS})$ (nb) & $B\cdot\sigma_{\rm NNLO}^Z(5{\rm FS})$ (nb)
  & $B\cdot\sigma_{\rm NNLO}^Z(5{\rm FS},b)$ (nb)\\  \hline
$\sigma_0^Z$   & 0.8083 & 0.8266 & 0.0202\\
$\sigma_1^Z$   & 0.1239 & 0.1322 & -0.0020\\
$\sigma_2^Z$   & 0.0037 & -0.0002 & -0.0037\\ \hline
total & 0.9359  & 0.9586 & 0.0145\\ \hline
$\Delta_b\sigma^Z$ & 0.0066  & $-$ & \\ \hline
total + $\Delta_b\sigma^Z$  & 0.9426  & 0.9586 &  \\
    \hline\hline\multicolumn{4}{c}{}\\\hline\hline
    LHC, $\sqrt{s} = 14$ TeV &  $B\cdot\sigma_{\rm NNLO}^Z(4{\rm FS})$ (nb) & $B\cdot\sigma_{\rm NNLO}^Z(5{\rm FS})$ (nb)
  & $B\cdot\sigma_{\rm NNLO}^Z(5{\rm FS},b)$ (nb)\\  \hline
$\sigma_0^Z$   & 1.7472 & 1.8110 & 0.0641\\
$\sigma_1^Z$   & 0.2384 & 0.2557 & -0.0050\\
$\sigma_2^Z$   & -0.0047 & -0.0153 & -0.0107\\ \hline
total & 1.9809  & 2.0514 & 0.0484\\ \hline
$\Delta_b\sigma^Z$ & 0.0231  & $-$ & \\ \hline
total + $\Delta_b\sigma^Z$  & 2.0040  & 2.0514 & \\
\hline\hline
  \end{tabular}
  \caption{\sf NNLO predictions for the total $Z$ cross section (multiplied by leptonic branching ratio $B$) at the Tevatron and LHC using MSTW~2008 NNLO PDFs~\cite{Martin:2009iq} as input, broken down into the $\alpha_S^n$ ($n=0,1,2$) contributions, with $\{q=u,d,s,c;\;\alpha_S^{(4)};\;$4-flavour MSTW~2008~NNLO PDFs$\}$ in the 4FS calculation and $\{q=u,d,s,c,b;\; \alpha_S^{(5)};\;$5-flavour MSTW~2008~NNLO PDFs$\}$ in the 5FS calculation.  The final column gives the contribution to the 5FS cross sections from processes where the $Z$ couples directly to $b$ quarks.  The additional ${\cal O}(\alpha_S^2)$ contributions to the cross section arising from real and virtual $b$-quark processes, taken from Table~\ref{tab:ztotb}, are added to the 4FS cross section in the last line of each sub-table.}
  \label{tab:z012nnlotot}
\end{table}
In Table~\ref{tab:z012nnlotot} we add these real and virtual $b$-quark contributions to the bulk of the NNLO cross section coming from 4FS light quarks and gluons, and compare the total with the benchmark 5FS NNLO results presented in Ref.~\cite{Martin:2009iq}, in which the $b$ quark is treated as a massless parton in the subprocess cross sections, i.e.~both in the initial and final states and in loop contributions.  Also shown in Table~\ref{tab:z012nnlotot} are the ($b\bar b, bg, \ldots$) contributions in which the $Z$ couples directly to a $b$ quark and in which there is at least one $b$ or $\bar b$ quark in the initial state.\footnote{This excludes small ${\cal O}(\alpha_S^2)$ contributions initiated by light quarks and gluons, e.g.~$q \bar q, gg \to Z b\bar b$, in which the $Z$ couples to $b$ quarks.} These represent $0.03\%$, $1.5\%$ and $2.4\%$ of the total 5FS NNLO cross sections at the Tevatron, LHC(7 TeV) and LHC(14 TeV) respectively.

From Table~\ref{tab:z012nnlotot}, we see that at the Tevatron the 5FS and 4FS$ + b$-quark total cross sections are the same to within 1$\%$.  At the LHC, however, the additional $b$-quark contributions to the 4FS cross section are at the $+1\%$ level and so do not completely compensate the ``missing'' $b \bar b \to Z$ contributions to the 5FS cross section.  To be precise,
\begin{equation}
  B\cdot\sigma_{\rm NNLO}^Z(5{\rm FS},b) = 48.4\; (14.5) \ {\rm pb} > \Delta_b\sigma^Z = 23.1\; (6.6)\ {\rm pb}
  \label{eq:45diff}
\end{equation}
at $\sqrt{s} = 14$~TeV (7~TeV), which results in the ``full'' 4FS total cross section being $2.3\%$ ($1.7\%$) smaller than the the 5FS cross section.  We interpret this as meaning that the DGLAP-resummed $[\alpha_S \ln(M_Z^2/m_b^2)]^n$ contributions that are absorbed into the $b$ parton distribution are numerically important. 

Our results are in broad agreement with the study of Ref.~\cite{Maltoni:2005wd}, in which the different calculational methods for heavy particles produced in association with $b$ quarks, $gg\to Xb\bar b$ (4FS) versus $b \bar b \to X$ (5FS), were studied in detail for $X=H,Z$. In particular, it was shown that with the canonical choice of scale $\mu=M_Z$, the LO $gg\to Zb \bar b$ cross section at the LHC was a factor of two smaller than the NNLO $b \bar b \to Z$ cross section, consistent with our results in Eq.~(\ref{eq:45diff}) above.  It was also shown in Ref.~\cite{Maltoni:2005wd} that the agreement between the 4FS and 5FS calculations improves if the scale $\mu$ is reduced to around $M_Z/3$ (found by choosing $\mu\sim \sqrt{-t}$ near the end of the collinear plateau in the quantity $-t\,{\rm d}\sigma/{\rm d}t$ for the process $gb\to Zb$).  The NNLO $b \bar b \to Z$ cross section is approximately scale independent, while the LO $gg\to Zb \bar b$ cross section increases with decreasing scale, primarily because of the overall $\alpha_S^2$ factor.

Of course the explicit $gg \to Z b \bar b$ contribution corresponds only to the $n=1$ term in the resummed $[\alpha_S \ln(M_Z^2/m_b^2)]^n$ perturbation series implicit in the $b\bar{b}\to Z$ 5FS calculation. Complete agreement between the two schemes {\it would} be obtained in a fully all-orders perturbative calculation.  Note that since at all collider energies $\Delta_b\sigma^Z$ is dominated by the contributions involving the $Zb \bar b$ final state, we would expect that higher-order corrections to the $Z b \bar b$ production process, which is here calculated only at leading order, will generate the $[\alpha_S \ln(M_Z^2/m_b^2)]^n$ terms implicit in the  $b$-quark PDF.  The NLO (i.e.~${\cal O}(\alpha_S^3)$) corrections to the $Z b \bar b$ total cross sections with $m_b \neq 0$ have recently been calculated~\cite{FebresCordero:2006sj,FebresCordero:2008ci,Cordero:2009kv,zbbCMS}.  Although the results presented in Ref.~\cite{zbbCMS} impose a minimum $p_T^b$ ($>5$~GeV), the $K$-factor\footnote{The $K$-factor is here defined as the ratio of  cross sections calculated using the dynamical scale $\mu^2 = M_Z^2 + (p_T^{b,1})^2 + (p_T^{b,2})^2$, and with the standard 5-flavour CTEQ6M/CTEQ6L1 PDFs~\cite{Pumplin:2002vw} at NLO/LO.} is evidently rather independent of $p_T^b$ at small $p_T^b$, suggesting that the NLO/LO $K$-factor for the fully inclusive $Zb\bar b$ cross section is approximately 1.5 for the LHC at $\sqrt{s} = 14$~TeV. It is therefore plausible that even higher-order perturbative corrections can account for the factor of 2 difference in the 4FS and 5FS cross sections, Eq.~(\ref{eq:45diff}).  This conclusion is supported by the fact that the scale dependence of the 4FS calculation for $Zb\bar b$ production at NLO is only mildly weaker than at LO~\cite{FebresCordero:2006sj,FebresCordero:2008ci,Cordero:2009kv,zbbCMS}.

In Ref.~\cite{Thorne:1997ga} it was shown that for $x$ in the region of 0.01--0.05, the relevant region for $Z+b\bar{b}$ production at the 14~TeV LHC, the ratio of the GM-VFNS structure function $F_2^h$ to the FFNS structure function was $\sim 1.5$ at LO at high scales.  This represents the effect of either resumming the $[\alpha_S\ln(Q^2/m_h^2)]^n$ contributions, or keeping only the contribution of fixed order in $\alpha_S$ for one parton. For hadron--hadron processes we would expect the difference to be about $1.5^2>2$, exactly as observed.  At NLO for structure functions at this $x$ the ratio is reduced to $\sim1.1$, so inclusion of the extra $\ln(Q^2/m_h^2)$ removes much of the discrepancy present at LO. However, for hadron--hadron processes, NLO in the fixed flavour scheme only contains the extra large logarithm for one of the two incoming partons, so the ratio between the 5FS and 4FS would be roughly $1.5\times 1.1 \approx 1.6$, again as we expect to see in practice. It would only be at NNLO in the 4FS, when the double logarithm for both incoming PDFs is included, that we would expect to see the reduction to roughly $1.1^2 \approx 1.2$ in the ratio of the 5FS to 4FS cross sections. This is a general feature, i.e.~the 4FS (or 3FS) will converge to the resummed 5FS results more slowly for hadronic processes than for those in DIS.

In summary, the 5FS PDFs are clearly the most appropriate to use for inclusive quantities such as $\sigma^Z$ (or $\sigma^{\rm dijet}$, etc.) at high $Q^2$ scales,  where resummation of the $[\alpha_S \ln(Q^2/m_b^2)]^n$ contributions is evidently important.  However, for more exclusive quantities like $\sigma^{Zb\bar b}$, where the $b$ quarks are measured in the final state, the 4FS parton distributions are more appropriate since the 5FS calculation gives no information on the spectator $b$ and $\bar b$ quarks which must be present in the final state for the $b$- and $\bar{b}$-initiated processes.  Note that if only the \emph{total} cross section is required, without cuts imposed on the $b$-quarks, then a 5FS is still better, e.g.~for $Zb\bar{b}$ a 5FS calculation can be used for $b\bar{b}\to Z$ at NNLO, where the $b$-quarks couple directly to the $Z$, and so there are implicitly also two $b$-quarks in the final state~\cite{Maltoni:2005wd}.  However, if cuts must be applied to the $b$-quarks, as is the case in the experimental measurement, then a 4FS calculation is more appropriate.  Similar remarks apply to the calculation of $Hb\bar{b}$ production~\cite{Dittmaier:2003ej,Dawson:2003kb,Campbell:2004pu} and other processes where $b$-quarks are detected in the final state.  In a recent study~\cite{Dittmaier:2009np}, the production of a charged Higgs boson in association with a $t$ quark was considered in both the 4FS ($gg \to t \bar b H^-$ etc.) and 5FS ($g b \to t H^-$ etc.) to NLO in pQCD, using the appropriate MRST 2004 PDF sets~\cite{Martin:2006qz,Martin:2004ir}. The central predictions in the 5FS were shown to be approximately 40$\%$ larger than those in the 4FS. Even taking the scale uncertainty into account the 4FS and 5FS NLO cross sections are barely consistent.

An ideal calculation would combine the best features of the 4FS and 5FS so as to resum $[\alpha_S\ln(Q^2/m_b^2)]^n$ terms while also retaining a finite $m_b$ dependence in the partonic cross section (rather than setting $m_b$ to zero as done in the 5FS).  This matching has, of course, been done for structure functions in DIS using different variants of the GM-VFNS, but applications to hadron collider cross sections are more involved and have so far been limited (see Ref.~\cite{Cacciari:1998it} for an application of the GM-VFNS to the $p_T$ spectrum in heavy-flavour hadroproduction).  However, for processes where the hard scale is, for example, $Q^2\sim M_Z^2$, then the GM-VFNS calculation will differ from the ZM-VFNS (5FS) only by terms $\mathcal{O}(m_b^2/M_Z^2\sim 0.3\%)$.  We would therefore expect the complete GM-VFNS calculation to give results very close to the pure ZM-VFNS (5FS) for the total cross section.

Note that rather than producing separate 4-flavour PDFs for use in a 4FS calculation, an alternative approach (e.g.~used at NLO in Refs.~\cite{Campbell:2009ss,Campbell:2009gj}) is to use the conventional 5-flavour PDFs, then pass to the 4-flavour scheme using counterterms given in Ref.~\cite{Cacciari:1998it}.  However, these counterterms~\cite{Cacciari:1998it} are equivalent to using the inverse of transition matrix elements, but only out to order $\alpha_S$.  One could indeed use the transition matrix elements themselves to go from 4-flavour to 5-flavour PDFs, except this would not sum the logarithmic terms, $[\alpha_S \ln(Q^2/m_b^2)]^n$, in the PDF evolution.\footnote{See Fig.~9 of Ref.~\cite{Alekhin:2009ni} for a comparison of 5-flavour NNLO PDFs obtained from 3-flavour PDFs either by evolution or by applying fixed-order matching conditions; the differences will be larger at NLO.}   Hence, the use of counterterms~\cite{Cacciari:1998it} is a less complete way of going from a 5FS to a 4FS, and instead, we recommend that dedicated 4-flavour PDFs be used in 4FS calculations.\footnote{However, for the 4FS calculation of $t$-channel single-top production at the Tevatron and LHC~\cite{Campbell:2009ss}, it was explicitly checked that results obtained with the dedicated 4-flavour MRST set~\cite{Martin:2006qz} were consistent (within the numerical integration precision) with those obtained with the corresponding 5-flavour MRST set~\cite{Martin:2004ir} plus appropriate counterterms~\cite{Cacciari:1998it}.  We thank R.~Frederix and F.~Tramontano for discussions on this issue.}  Previously, a major advantage of using 5-flavour PDFs with counterterms was that eigenvector PDF sets to calculate PDF uncertainties were not made available for existing 4-flavour PDFs~\cite{Martin:2006qz}.  However, we have now provided eigenvector PDF sets also for the 4-flavour PDFs, therefore this advantage no longer holds.

\section{Conclusions} \label{sec:conclusions}

We have repeated the NLO and NNLO MSTW~2008 global PDF analyses~\cite{Martin:2009iq} for a range of heavy-quark masses about their ``default'' values $m_c=1.40$~GeV and $m_b=4.75$~GeV.  For the charm quark, we found that the global data prefer the values $m_c = 1.45$ (1.26)~GeV at NLO (NNLO).  The most discriminating data are, as anticipated, the HERA data for $F_2^c$~\cite{Adloff:1996xq,Adloff:2001zj,Aktas:2005iw,Aktas:2004az,Breitweg:1999ad,Chekanov:2003rb,Chekanov:2007ch}.  On the other hand, for the bottom quark, the data included in the global fit (excluding $F_2^b$) do not put a meaningful constraint on the value of $m_b$, while the HERA $F_2^b$ data slightly favour $m_b \approx 4.75$--5~GeV.  We pointed out that precise determinations of the heavy-quark masses in the $\overline{\rm MS}$ scheme are affected by poorly convergent perturbative series in the conversion to the pole masses, particularly for the case of the charm quark.  Recent precise combined HERA data on $\sigma_r^{\rm NC}$~\cite{Aaron:2009wt} and $F_2^c$~\cite{HERAF2charm} will in future be able to narrow the favoured range of the charm-quark pole mass $m_c$.  Note, however, that uncertainties from the choice of GM-VFNS~\cite{Thorne:2010pa} mean that the favoured value of $m_c$ will be correlated to some extent with the particular choice of GM-VFNS, although this correlation will be much smaller at NNLO than at NLO, as will other uncertainties arising from the choice of GM-VFNS~\cite{Thorne:2010pa}.

We explored the effect of the values of the heavy-quark masses on $W$, $Z$ and Higgs production at the Tevatron and LHC.  Varying the charm mass by $\pm 0.15$ GeV changes the cross sections by  $\pm 1\%$ or less at Tevatron energies and by $\pm 2\%$ at the LHC energy of $\sqrt{s}=14$ TeV.  The various weak boson cross-section \emph{ratios} are essentially unchanged. The predictions for $W$, $Z$ and Higgs cross sections are much less dependent on the value taken for $m_b$.  We provided a recommendation on how to include the uncertainty arising from the choices of $m_c$ and $m_b$ in a generic cross-section calculation.

We also presented PDF sets obtained in a framework with different active numbers of quark flavours, as done previously in the context of the MRST~2004 analysis~\cite{Martin:2006qz}.  Explicitly, we determined 4-flavour PDF sets in which charm becomes an active parton, but bottom does not, and 3-flavour PDF sets where charm and bottom are not partons, but only appear in the final state.  The analogous 5-flavour parton sets are simply those of MSTW~2008~\cite{Martin:2009iq}.  Of course, the latter, which in the absence of top corresponds to PDFs of a (general-mass) variable flavour number scheme, are generally to be preferred, particularly for inclusive quantities at high $Q^2$ scales where the resummation of the $[\alpha_S\ln(Q^2/m_b^2)]^n$ contributions is essential.  However, for more exclusive processes, such as $Zb\bar{b},~Hb\bar{b},~\ldots$, where $b$-quarks are observed in the final state, the 4-flavour parton distributions are more appropriate.  For illustration, we computed the various components of the $Z$ production cross section to ${\cal O}(\alpha_S^2)$ at the Tevatron and LHC, and compared the predictions obtained using the 4FS and 5FS ($\equiv$ MSTW~2008~\cite{Martin:2009iq}) parton sets.

The additional grids for all PDF sets discussed in this paper are made publicly available~\cite{mstwpdf}, for use either with the standalone MSTW interpolation code or via the \textsc{lhapdf} interface~\cite{Whalley:2005nh}.  To be precise, grids for the following PDF sets are made available:
\begin{itemize}
  \item For the default quark masses, $m_c=1.40$~GeV and $m_b=4.75$~GeV, we provide LO, NLO and NNLO grids for 3- and 4-flavour PDFs (central set and 40 eigenvector sets at both 68\% and 90\% C.L.).  These grids complement the existing grids for the 5-flavour PDFs.
  \item For $m_c$ in the range 1.05~GeV to 1.75~GeV (in steps of 0.05~GeV), we provide NLO and NNLO grids for 3- and 5-flavour PDFs (central set only) for both free and fixed $\alpha_S(M_Z^2)$.
  \item For $m_b$ in the range 4.00~GeV to 5.50~GeV (in steps of 0.25~GeV), we provide NLO and NNLO grids for 4- and 5-flavour PDFs (central set only) for free $\alpha_S(M_Z^2)$ only.
\end{itemize}
These additional grids should prove to be useful in future for detailed studies of a variety of collider processes involving heavy quarks.

\appendix
\setcounter{equation}{0}
\renewcommand{\theequation}{A.\arabic{equation}}
\section{Appendix} \label{sec:appendix}

We consider the $b$-quark contributions to the 4FS calculation, closely following the discussion in Ref.~\cite{Rijken:1995gi}. First, we have the two-loop corrections to the $q + \bar q \to Z$ Born process, examples of which are shown in Fig.~\ref{fig:diagrams}(a) and (b).  In the notation of Ref.~\cite{Rijken:1995gi},
\begin{equation}
W_{q\bar q}^Z = \delta(1-{\hat\tau}) C_F T_f \left( \frac{\alpha_S}{\pi}\right)^2 \left[ {2}a_q a_Q
G_1(\rho_b)  + \frac{1}{2}\left( v_q^2 + a_q^2  \right) F(\rho_b) \right] ,
\label{eq:1a1b}
\end{equation}
with $\rho_b = m_b^2/M_Z^2 \approx 2.7 \times 10^{-3}$. The functions $F$ and $G_1$ are defined in Refs.~\cite{Rijken:1995gi} and~\cite{Gonsalves:1991qn} respectively. Since in practice $\rho_b \ll 1$, we can use the following small-mass expansions:
\begin{eqnarray}
F(\rho) &\simeq&  -\frac{4}{9}L^3 + \frac{38}{9}L^2 + \left( \frac{16}{3}\zeta_2  -\frac{530}{27}\right) L 
+ \frac{3355}{81} - \frac{152}{9}\zeta_2 -\frac{16}{3}\zeta_3  \nonumber \\
& & +  \rho\left( 16 L^2 - 32 L + 128 -\frac{192}{3}\zeta_2  \right)  + \mathcal{O}(\rho^2) ,\\
G_1(\rho) &\simeq&  4\rho\left( -2 \zeta_2 L + 2 \zeta_3 - 3\right) + \mathcal{O}(\rho^2) ,
\end{eqnarray}
with $L = \ln(1/\rho) \approx 5.9$ for $\rho = \rho_b$. The contributions to the total $Z$ cross section from these loop diagrams are given in the first row of Table~\ref{tab:ztotb}. They are numerically very small and negative (positive) at the LHC (Tevatron).\footnote{Even though the function $G_1(\rho)$ is suppressed by a power of $\rho$ relative to $F(\rho)$ at small $\rho$, it gives a comparable contribution to the cross section. This is because $F$ has a zero close to the physical value $\rho = \rho_b$.}

Next we have the one-loop corrections to the $2\to 2$ processes $q + \bar q \to Z + g$ and $g + q(\bar q) \to Z + q(\bar q)$, examples of which are shown in Fig.~\ref{fig:diagrams}(c) and (d):
\begin{equation}
W_{q\bar q}^Z = \frac{1}{2}a_q a_Q C_F T_f \left( \frac{\alpha_S}{\pi}\right)^2 
\left[ \frac{1+{\hat\tau}}{1-{\hat\tau}}\left\{-2 + 2{\hat\tau} \left(J_1(4\rho_b{\hat\tau}) -J_1(4\rho_b)\right)\right\}
- 4 \rho_b{\hat\tau}  \left(J_2(4\rho_b{\hat\tau}) -J_2(4\rho_b)\right) \right]  ,
\label{eq:2a}
\end{equation}
\begin{equation}
W_{q g}^Z = \frac{1}{2}a_q a_Q T_f^2 \left( \frac{\alpha_S}{\pi}\right)^2 
H_1({\hat\tau} ,\rho_b) ,
\label{eq:2b}
\end{equation}
where $\hat s = {\hat\tau}^{-1} M_Z^2$ and the functions $H_1$, $J_1$ and $J_2$ are defined in Ref.~\cite{Gonsalves:1991qn}. The following small-mass approximations are again useful:
\begin{eqnarray}
H_1(\tau , 0)  & = & 2\tau \left[ \left\{ 2(\tau -1) + \ln\left( \frac{1}{\tau} \right) \right\} \ln\left( \frac{1-\tau}{\tau} \right) + {\rm Li}_2\left( \frac{\tau - 1}{\tau} \right) \right] + 2(1-\tau) , \nonumber \\
J_1(\rho) &\simeq&  L + \ln4 -2 -\frac{\rho}{2}\left( L +\ln4 +1 \right)  + \mathcal{O}(\rho^2) ,  \nonumber \\
J_2(\rho) &\simeq&  \frac{1}{2}\left( (L + \ln4)^2 -\pi^2 \right)  -\frac{\rho}{2}\left( L +\ln4 \right)  + \mathcal{O}(\rho^2) .
\label{eq:2c}
\end{eqnarray}

Finally, we have the $2\to 3$ processes
\begin{equation}
q + \bar q \to Z + b + \bar b, \quad
g +      g \to Z + b + \bar b ,
\label{eq:3}
\end{equation}
see Fig.~\ref{fig:diagrams}(e) and (f).  The calculation of these is in principle straightforward, with matrix elements squared integrated over three-body phase space. The non-zero $b$-quark mass makes these contributions infra-red and collinear finite, i.e.~the singularities present for massless quarks are replaced by terms with logarithmic ($\ln(M_Z^2/m_b^2)$) behaviour in the small $m_b$ limit.  We use the \textsc{mcfm} implementation~\cite{Campbell:2000bg} of the matrix elements and phase space, with exactly the same parameter choice as in the previous MSTW calculations.  The results are given in Table~\ref{tab:ztotb}.  Note that the $gg$ contribution is larger than the $q \bar q$ contribution at the LHC, while the converse is true at the Tevatron.

\end{document}